\documentclass{article} 
\usepackage{graphicx}
\usepackage{fullpage}
\usepackage{amsmath}
\usepackage{url}
\usepackage{gensymb}
\usepackage{changepage}
\usepackage{setspace}
\usepackage{booktabs}
\usepackage{color}
\usepackage{multirow}
\usepackage{enumitem}
\usepackage{caption}
\usepackage{subcaption}
\usepackage{array}
\usepackage{multicol}
\usepackage{lipsum}
\usepackage{longtable}
\usepackage{float}

\usepackage[tableposition=top]{caption}
\restylefloat{table}
\usepackage{authblk}
\usepackage[normalem]{ulem}
\usepackage[table]{xcolor}
\definecolor{coolblack}{rgb}{0.0, 0.18, 0.39}
\definecolor{applegreen}{rgb}{0.55, 0.71, 0.0}
\definecolor{mediumcarmine}{rgb}{0.69, 0.25, 0.21}
\definecolor{mediumlavendermagenta}{rgb}{0.8, 0.6, 0.8}
\definecolor{selectiveyellow}{rgb}{1.0, 0.73, 0.0}
\definecolor{blue}{rgb}{0,0,1} 

\title{More than one Author with different Affiliations}
\author[a]{R. Radhakrishnan}
\author[a]{W. W. Edmonson}
\author[b]{F. Afghah}
\author[c]{R. M. Rodriguez-Osorio}
\author[a]{F. Pinto Jr.}
\author[d]{Scott C. Burleigh}
\affil[a]{Department of Electrical and Computer Engineering,
 North Carolina A\&T State University, North Carolina, USA}
 \affil[a]{\textit {\{rradhakr, wwedmons, fmpinto\}@ncat.edu}}
\affil[b]{Department of Electrical Engineering and Computer Science, Northern Arizona University, Flagstaff, Arizona, USA}
 \affil[b]{\textit {\{fatemeh.afghah\}@nau.edu}} 
\affil[c]{Polytechnial  University of Madrid, Madrid, Spain}
 \affil[c]{\textit {\{ramon\}@gr.ssr.upm.es}}
\affil[d]{Jet Propulsion Laboratory, California Institute of Technology, Pasadena, California, USA}
\affil[d]{\textit {\{Scott.C.Burleigh\}@jpl.nasa.gov}}
\date{}
\begin{document}

\title{Survey of Inter-satellite Communication for Small Satellite Systems: Physical Layer to Network Layer View}
\maketitle

\begin{abstract}
Small satellite systems enable whole new class of missions for navigation, communications, remote sensing and scientific research for both civilian and military purposes. As individual spacecraft are limited by the size, mass and power constraints, mass-produced small satellites in large constellations or clusters could be useful in many science missions such as gravity mapping, tracking of forest fires, finding water resources, etc. The proliferation of small satellites will enable a better understanding of the near-Earth environment and provide an efficient and economical means to access the space through the use of multi-satellite solution. Constellation of satellites provide improved spatial and temporal resolution of the target. Small satellite constellations contribute innovative applications by replacing a single asset with several very capable spacecraft which opens the door to new applications. Future space missions are envisioned to become more complex and operate farther from Earth which will need to support autonomous operations with minimal human intervention. With increasing levels of autonomy, there will be a need for remote communication networks to enable communication between spacecraft. These space based networks will need to configure and maintain dynamic routes, manage intermediate nodes, and reconfigure themselves to achieve mission objectives. Hence, inter-satellite communication is a key aspect when satellites fly in formation. In this paper, we present the various researches being conducted in the small satellite community for implementing inter-satellite communications based on the Open System Interconnection (OSI) model. This paper also reviews the various design parameters applicable to the first three layers of the OSI model, i.e., physical, data link and network layer. Based on the survey, we also present a comprehensive list of design parameters useful for achieving inter-satellite communications for multiple small satellite missions. Specific topics include proposed solutions for some of the challenges faced by small satellite systems, enabling operations using a network of small satellites, and some examples of small satellite missions involving formation flying aspects.
\end{abstract}

\section{Introduction}

In recent years, there is a growing interest in small spacecraft for missions in and beyond Lower Earth Orbit (LEO) particularly in the pico, nano, and micro class of satellites. Small satellites are artificial satellites with lower weights and smaller sizes and is becoming more attractive due to lower development costs and shorter lead times~\cite{NASAreport01}. Small satellites, usually under 500 Kg, are classified according to their mass into mini-satellite, micro-satellite, nano-satellite (cube satellite), pico-satellite and femto-satellite~\cite{NASAreport01}, as shown in Table~\ref{tab:tab1}. There are numerous constraints for small satellites because of size, power, and mass. However, miniaturization and integration technologies have diminished the trade-off between size and functionality. These classes of satellites enable missions that cannot be accomplished by large satellites such as high temporal and spatial resolution by gathering data from multiple points, in-orbit inspection of large satellites, ease of mass production, space missions consisting of large number of satellites forming constellations or loose clusters, and university related research~\cite{Wikismallsat}.\\
\begin{table}[h!]
\caption{Small spacecraft classifications}
\label{tab:tab1}

\begin{center}
\begin{tabular}{| >{\centering\arraybackslash}m{4 cm} | >{\centering\arraybackslash}m{3 cm} |}
\hline
\textbf{Type of satellite} & \textbf{Mass} \\
\hline
Mini-satellite & 500-100 Kg \\
\hline
Micro-satellite & 100-10 Kg\\
\hline
Nano-satellite (cubesat) & 10-1 Kg  \\
\hline
Femto and Pico-satellite & $<$ 1 Kg \\
\hline
\end{tabular}
\end{center}
\end{table}

\begin{figure}[h]
 \centering
 \renewcommand{\figurename}{Figure}
 \includegraphics[width =3.2in, height = 1.3in]{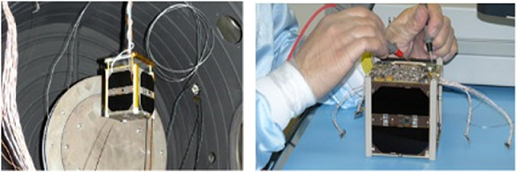}
 \caption{The Swiss Cube~\cite{swisscube}}
  \label{fig:Swiss cube 01}
 \end{figure}

Small satellites serve as a platform for the development of new space technologies, allowing non-spacefaring nations, companies, universities, scientists, and engineers all over the world to have low cost access to space. There are several companies or organizations that design, manufacture, and launch advanced rockets and spacecraft, such as SpaceX~\cite{SpaceX}, Orbital Sciences Cooperation~\cite{orbitalscience}, NanoRacks~\cite{nanoracks}, Planet Labs~\cite{planetlabs}, Skybox~\cite{skybox}, Pumpkin~\cite{pumpkin}, etc. Total launch cost for small satellites are under a few million dollars in comparison to \$200-1000 million for a full-sized one. The Boeing launch vehicle aimed to launch small payloads of 45 Kg, with cost as low as \$300,000 per launch, using their Small Launch Vehicle (SLV) concept, which could be in service by 2020~\cite{Boeing}. The minimum price of a pico-satellite (the size of a soda can) launch is \$12,000~\cite{DIY}. Over the last 50 years, more than 860 micro-satellites, 680 nano-satellites, and 38 pico-satellites have been launched globally~\cite{smallsatbook}. Figure~\ref{fig:Swiss cube 01} shows an example of a small satellite, the Swiss Cube, developed by 'Ecole polytechnique F\'ed\'erale de Lausanne (EPFL)'s space center, which is still in operation (as of February 2014). It was launched on September 23, 2009, for a mission duration of three months to one year. The mission aimed to photograph ``air glow", a phenomenon occurring due to the interaction between solar radiation and oxygen molecules in the upper atmosphere~\cite{swisscube}. \\
 
 A large number of heterogeneous small satellites can be deployed in space as a network using inter-satellite communications to enable command, control, communication and information processing with real time or near real time communication capabilities. The concept of multiple satellite mission is becoming attractive because of their potential to perform coordinated measurements of remote space, which can also be classified as a sensor network. Multi-satellite solution is highly economical and helps to provide improved spatial and temporal resolutions of the target. A large number of heterogeneous small satellites can be deployed in space as a network with minimum human intervention, and thus demanding a need for Inter-Satellite Communications (ISC). Future space missions requiring Distributed Space Systems (DSS) will consist of multiple advanced, intelligent and yet affordable satellites in space that communicates with each other, which could enable an unprecedented amount of communications and computing capabilities from which the satellite industry, university researchers, and scientists all over the world could benefit.\\
 
The novelty of this survey paper is that this is the first work to summarize the various research being conducted in the area of inter-satellite communications for small satellites and to provide a complete architecture based on the Open System Interconnection (OSI) model framework for small satellite networks. This article surveys the literature over the period 2000-2015 on inter-satellite communications as they apply to small satellites. The paper provides an extensive survey of significant number of design approaches of various layers of the OSI model for small satellite systems, in particular the first three layers, i.e., physical, data link, and network layer. The upper layers of the OSI model which are application/mission-specific, are beyond the scope of this paper.”

\subsection{Paper Organization}
The paper is organized in the following manner. A list of all abbreviations used in this paper is given in Section 1.2. A brief overview of the various applications of small satellites is given in Section 1.3. Section 1.4 provides some examples of launched/proposed small
satellite missions involving large number of satellites with inter-satellite communications. Section 2 serves as an introduction to various configurations of small satellites and motivation for inter-satellite communications in multiple small satellite missions. Section 3 presents an extensive survey on the various design approaches for inter-satellite communications in small satellite systems, in terms of the first three layers of the OSI model, i.e., physical, data link, and network layer. This paper also presents solutions to some of the challenges faced by small satellite systems including interference mitigation using hybrid multiple access protocol~\cite{ISC07} and software defined radio implementation for inter-satellite communication~\cite{VTCconf} in Section 4. Section 5 suggests specific directives to consider to the readers for the design and development of various parameters of the OSI model for multiple small satellite systems. Future research directions are illustrated in Section 6 and the paper is concluded in Section 7.
\subsection{List of Abbreviations}
Table~\ref{tab:tabnew} shows the list of abbreviations used in this paper.\\
\begin{table}[H]
\caption{List of abbreviations}
\label{tab:tabnew}
\begin{center}
\begin{tabular}{|c|c|}
\hline
\textbf{Acronym} & \textbf{Abbreviation} \\
\hline
AOCS & Attitude and Orbit Control System\\
\hline
BDSR & Bandwidth Delay Satellite Routing \\
\hline
BP & Bundle Protocol\\
\hline
BPSK & Binary Phase Shift Keying\\
\hline
BTMA & Busy Tone Multiple Access\\
\hline
CCSDS & Consultative Committee for Space Data Systems\\
\hline
CDMA & Code Division Multiple Access\\
\hline
CSMA & Carrier Sense Multiple Access\\
\hline
CTS & Clear-to-Send\\
\hline
CW & Contention Window\\
\hline
DIFS & Distributed co-ordination function Inter-Frame Space\\
\hline
DSS &  Distributed Space Systems\\
\hline
DTN & Delay Tolerant Networking\\
\hline
EIFS & Extended Inter-Frame Space\\ 
\hline
FDD & Frequency Division Duplex\\
\hline
FDMA &  Frequency Division Multiple Access\\
\hline
FSK & Frequency Shift Keying\\
\hline
GNSS & Global Navigation Satellite System\\
\hline
GPS & Global Positioning System\\
\hline
IF & Intermediate Frequency\\
\hline
ISC & Inter-Satellite Communications\\
\hline
ISM & Industrial, Scientific, and Medical\\
\hline
ISMA & Idle Signal Multiple Access\\
\hline
LDPC &  Low Density Parity Check \\
\hline
LEO & Lower Earth Orbit\\
\hline
MAC & Medium Access Control\\
\hline
MAI & Multiple Access Interference\\
\hline
MDR & Maximum Data Rate\\
\hline
OBDH & On-Board Data Handling\\
\hline
OFDM & Orthogonal Frequency Division Multiple Access\\
\hline
OLFAR & Orbiting Low Frequency Antennas for Radio Astronomy\\
\hline
OSI & Open System Interconnection\\
\hline
PFF & Precision Formation Flying\\
\hline
QPSK & Quadrature Phase Shift Keying\\
\hline
RTS &  Request-to-Send\\
\hline
SCA & Software Communication Architecture\\
\hline
SDMA & Space Division Multiple Access \\
\hline
SDR & Software Defined Radio\\
\hline
SIFS & Short Inter-Frame Space\\
\hline
SMAD & Space Mission Analysis and Design\\
\hline
TCP & Transmission Control Protocol\\
\hline
TDD & Time Division Duplex\\
\hline
TDMA & Time Division Multiple Access\\
\hline
UDP & User Datagram Protocol\\
\hline
UHF/VHF & Ultra/Very High Frequency\\
\hline
USRP & Universal Serial Radio Peripheral\\
\hline
WLAN & Wireless Local Area Network\\
\hline
\end{tabular}
\end{center}
\end{table}
In the next section, we provide an overview of the various applications using network of small satellites. 
\subsection{Enabling Operations Using Network of Small Satellites}
Network of small satellites should be capable of operating symbiotically. Examples of these type of operations are servicing or proximity operations, autonomous operations, fractionated spacecraft, and distributed processing. A more detailed description of these examples are described below.
\subsubsection{Servicing or proximity operations}

It is a new trend in research to assess the feasibility, practicality, and cost of servicing satellites and space stations using several spacecraft with robotic capabilities. There are numerous advantages of proximity operations: increasing the value of extremely useful assets (for e.g., international space station), removal of space debris, injection error occurred due to the malfunction of the upper stage of the launcher that could be corrected by an on-orbit servicing spacecraft, thus increasing the overall success rates of space missions, and also repairing and refueling of commercial satellites rather than replacing it~\cite{proximityref}. Servicing spacecraft could be used mainly in the geosynchronous orbits since it is extremely expensive to design, construct, and launch spacecraft in GEO orbits and hence, it is preferable to extend the lifetime of GEO spacecraft. Thus, the hardware and software components of these spacecraft, for proximity operations, should be capable of withstanding radiations and may require additional shielding. For deep space operations, the accurate location of the satellites cannot be obtained using GPS constellation. The relative location of the satellites and clock synchronization can be achieved using inter-satellite communications. The X-ray emitting pulsars provide the ability to autonomously determine the position anywhere in the solar system just as GPS does for Earth locations~\cite{NASAtechnologyroadmap}.   
\subsubsection{Autonomous operations} 
The space environment is dynamic and/or unpredictable, networking multiple spacecraft for  a heterogeneous system could be difficult, leading to delayed or disrupted communication links. In a centralized system, there could be scenarios when the master satellite loses its functionality or capability, thus requiring a new master satellite~\cite{TanyaAgent}. To solve these issues, new agent based computing platforms are proposed, i.e., the satellites should have capabilities to perform intelligent improvements based on the situation. Agents are high abstraction of programming for complex problems. It is always beneficial to design goal oriented agents. Agents should have two basic functionalities: perception, i.e., how the agent view its environment or how agent is aware about the situation; and cognition, i.e., the actions an agent needs to take at any given situation. Each satellite in the system receives information from the neighboring satellites and decides the actions it should perform among the set of actions and move to the next state. Satellites need to discover the current network topology they have formed and should determine whether that situation is appropriate to initiate communication. In other words, satellites should recognize all possible combinations of network topologies they may form and wisely decide a suitable one for communication, so that, an optimum system performance can be achieved. Part of this decision making process is the utility function associated with each action that the satellite can carry out.
\subsubsection{Fractionated spacecraft} 
A single spacecraft can be fractionated into several homogeneous or heterogeneous modules that communicate via wireless links forming a highly dynamic topology. The modules form a cluster with mobile architecture, where the modules may or may not join the cluster. If a sensor or software component fails, the cluster must reconfigure itself autonomously to achieve the mission objectives, i.e., the architecture must exhibit fault tolerance. The software system in the modules should be designed along these lines to meet the challenges introduced by the fractionated system architectures~\cite{fractionatedref, fractionated2014}.  
\subsubsection{Distributed processing}
Distributed processing refers to the decentralization of computing resources or processors which may be physically located in different components or subsystems rather than a single large system. These processors may have sharing capabilities with collaborative architecture focusing on a specific mission~\cite{distributedprocessingref}. A distributed computing system has various architectural configurations, for example, star, ring, linear bus, hybrid, layered, etc. Decentralization of computing capabilities offers numerous advantages: $1)$ each functional block can be designed with precision and transparency, specifying the task of each component and the information exchange needed to initiate the task, $2)$ it allows easy scaling of functional and data flow designs for multiple satellite missions and also space/ground segments, $3)$ it will promote meticulous test and verification of individual components during the design and development phase, $4)$ distributed architecture will simplify resource sharing among various subsystems, thereby promoting fault tolerant capabilities by supplying computational functionalities in the event of failures.
\subsection{Small Satellite Missions Involving Formation Flying Aspects } 
In this section, we introduce a brief review of recent small satellite missions proposed or launched by several organizations and space agencies applying formation flying concepts. Table~\ref{tab:missions} shows some of the important multiple small satellite missions which are designed and developed by various space agencies and organizations. 

\begin{table*}[t]
\caption{Multiple small satellite missions and its related information}
\label{tab:missions}
\renewcommand{\arraystretch}{1.3}


\begin{center}


\begin{tabular}{|>{\centering\arraybackslash}m{1.5cm} | >{\centering\arraybackslash}m{2cm} | >{\centering\arraybackslash}m{2.5cm}| >{\centering\arraybackslash}m{2.5cm} | >{\centering\arraybackslash}m{3.3cm} | >{\centering\arraybackslash}m{3.5cm} |}

\hline

\textbf{Mission name} & \textbf{Number of small satellites} & \textbf{Mass of small satellites (Kg)} & \textbf{Inter-satellite links } & \textbf{Inter-satellite communication approach} & \textbf{Launched/Projected launch year} \\

\hline

GRACE & 2 & 480 & Available & RF based (S-band) & 2002 \\

\hline
ESSAIM & 2 & 120 & Not available & Not available & 2004 \\
\hline
PRISMA & 4 & 145, 50 & Available & RF based (UHF-band)  & 2010 \\
\hline
ELISA & 4 & 130 & Not available & Not available & 2011 \\
\hline
EDSN  & 8 & 1.7 & Available & RF based (UHF-band)  & 2015 \\
\hline
QB-50 & 50 & 2, 3 & Available & RF based (S-band)  & 2016 \\
\hline
PROBA-3 & 2 & 320, 180 & Available & RF based (S-band) & 2017 \\
\hline
eLISA  & 3 & To be determined & Available & Optical based (LASER)  & 2028 \\
\hline
MAGNAS  & 28 & 210, 5 & Available & RF based (UHF-band)  & To be determined \\
\hline
\end{tabular}



\end{center}

\end{table*}
\begin{enumerate}[label=(\alph*)]
\item GRACE - The GRACE (Gravity Recovery and Climate Experiment) mission is a joint venture of  National Aeronautics and Space Administration (NASA) in the United States and Deutsche Forschungsanstalt für Luft und Raumfahrt (DLR) in Germany which was launched in 2002, with two satellites, each of 480 kg, separated by 220 km in a polar orbit 500 km above the Earth. The primary objective of the mission is to accurately map the variations in Earth's magnetic field. The telemetry tracking and command system is using S-band frequencies for uplink, downlink, and crosslink communications~\cite{grace01,grace02}. 
\item ESSAIM - It is a French military exploration satellite constellation, launched in December 2004, with four small satellites, each 120 kg, flying in formation in two out of phase polar orbits, maintained at a mean altitude of 658 km. The primary mission objective was to analyze the electromagnetic environment on the ground for a number of frequency bands used exclusively for military applications. At the receiver end, X-band terminals are used to receive the stored data from the satellites as they come in line of sight with the ground segment~\cite{essaim01,essaim02}.
\item PRISMA - Prototype Research Instruments and Space Mission technology Advancement (PRISMA) was designed and developed by Swedish Space Cooperation (SSC) to demonstrate formation flying and rendezvous technologies. It consists of two spacecraft, one advanced and highly maneuverable called MAIN (MANGO, 145 kg), and a smaller spacecraft without a maneuvering capability called TARGET (TANGO, 50 kg). The MAIN communicates in S-band for downlink  and uplink and the TARGET communicates its position and status with MAIN using an inter-satellite link in the UHF band~\cite{prisma}.
\item ELISA - It is a demonstration project for mapping the positions of radar and other transmitters around the world and analyzing their characteristics, and is sponsored by French defense procurements agency, launched by a Russian Soyuz launcher in 2011. The ELISA involves 4 micro-satellites, each of 130 kg, placed in sun synchronous orbit at an altitude of around 700 km which are separated by few kilometers from each other~\cite{elisa01,elisa02}. 
\item EDSN - Edison Demonstration of Smallsat Network (EDSN) is NASA's first project to demonstrate small satellite applications using consumer electronics-based nano-satellites, consisting of a swarm of 8 cube satellites, each of mass $\approx$ 1.7 Kg with a smart phone on-board (Nexus S). The communication subsystems will use UHF band for cross links at a data rate of 9.6 Kbit/s, and S-band to communicate with the ground station~\cite{edsn}. 
\item QB-50 - The QB-50 mission concept is developed by the Von Karman Institute and is funded by the European Union. The goal of the project is to have an international network of 50 double and triple cubesats in a string of pearl configuration, which will allow multi-point, in-situ, and long duration exploration of lower thermo sphere at an altitude of 90-380 km. The objectives of the mission are the in-orbit demonstration of multi-spacecraft for in-situ measurements and atmospheric research within the lower thermo sphere. The satellites use UHF/VHF band for uplink and downlink communications and the project is scheduled to launch in 2016~\cite{qb50}. 
\item PROBA-3 - Project for on-board autonomy-3 is a small satellite technology development and demonstration mission by European Space Agency (ESA) scheduled to launch in 2017 at altitude of 600 km. The primary objective of the mission is to demonstrate the technologies needed for formation flying of multiple spacecraft. The PROBA-3 mission consists of two spacecraft referred to as CSC (Coronograph SpaceCraft) with a mass of $\approx$ 320 kg and OSC (Occulter SpaceCraft) with a mass of $\approx$ 180 kg. Inter-satellite links will be established using an S-band system between the spacecraft and the relative position of the satellites, obtained from GPS receivers, will be propagated where GPS signals are not available~\cite{proba302}.
\item eLISA -  Evolved Laser Interferometer Space Antenna (eLISA) is the space mission concept designed by European Space Agency to detect and accurately measure gravitational waves. The mission consists of a constellation of three satellites (one ``Mother" and two``Daughter") deployed in three different orbits maintaining a near equilateral triangular formation. X-band links will be used for communication between the ``Mother" spacecraft and ground. It is expected to launch in 2028 and would be an ideal tool for better understanding of the universe~\cite{elisaesa01,elisaesa02}.
\item MAGNAS - The Magnetic Nano-Probe Swarm mission is a concept expanded on ESA's SWARM mission, using a constellation of several nano-satellites in order to acquire simultaneous measurements of the geomagnetic field resolving the local field gradients. The MAGNUS system comprises of 4 spacecraft swarms, with each swarm consisting of 6 nano probes and 1 mother spacecraft. Each mother spacecraft and nanoprobe have mass around 210 kg and 5 kg respectively. The mother spacecraft uses S-band frequency to communicate to the ground and UHF to nanoprobe~\cite{magnas01}. 
\end{enumerate}

\section{Background}

In this section, we provide a brief introduction of different configurations of small satellites, particularly, leader-follower, cluster, and constellation formation flying patterns and then we explain the importance of inter-satellite communications when small satellites are deployed as a network in space.
\subsection{Satellite Formation Flying}

Multiple small spacecraft provide higher efficiency gain by promoting adaptability, scalability, reconfigurability, and affordability compared to a single large satellite. When satellites fly in formation, it is required to maintain specific distance and orientation relative to each other at specified altitudes. Depending on the formation characteristics, there can be two different approaches: ground based control and autonomous operations~\cite{Satelliteformation}. In ground based control, formation flying satellites send navigational measurements to the ground control center that provides necessary instructions to maneuver into appropriate position in the formations. This approach is suitable for formations with several kilometers of separation distance between the satellites. In autonomous formation flying, measurements are transmitted among the spacecraft allowing the satellites to calculate the relative position in the formation and Attitude and Orbit Control System (AOCS) is used to maneuver the satellite into appropriate positions. Autonomous approach is more difficult and riskier and is suitable for missions that require tighter formations with frequent and autonomous adjustments of the relative positions.  \\

There are different types of formations depending on the separation between vehicles and intended applications. The three most common types of formations are: trailing or leader-follower, cluster, and constellation~\cite{Satelliteformation}. 
\begin{enumerate}[label=(\alph*)]
\item Trailing - In this type of formation, multiple spacecraft share the same orbit and they follow each other at a specific distance. Figure~\ref{fig:trailing} shows the trailing formation flying pattern. 
\begin{figure}[h]
\centering
\renewcommand{\figurename}{Figure}
\includegraphics[width = 3in, height = 2.5in]{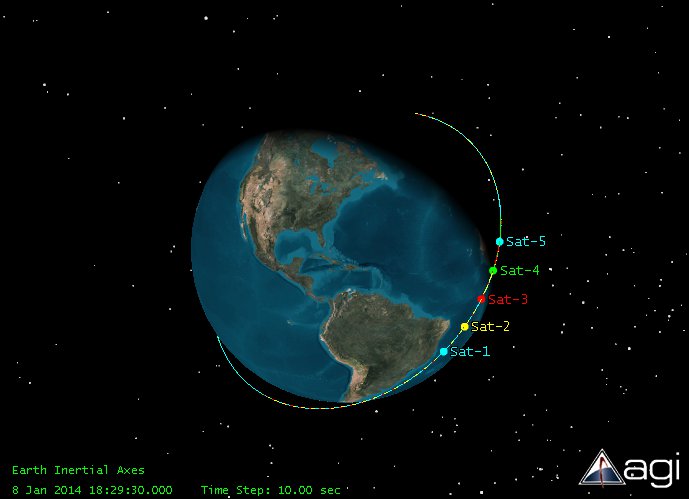}
\caption{Trailing formation flying pattern}
\label{fig:trailing}
\end{figure}
\item Cluster - A group of satellites will be deployed in their respective orbits and remain closer to each other covering a smaller portion of the Earth. Figure~\ref{fig:cluster} shows the cluster formation flying pattern. 
\begin{figure}[h]
\centering
\renewcommand{\figurename}{Figure}
\includegraphics[width = 3in, height = 2.5in]{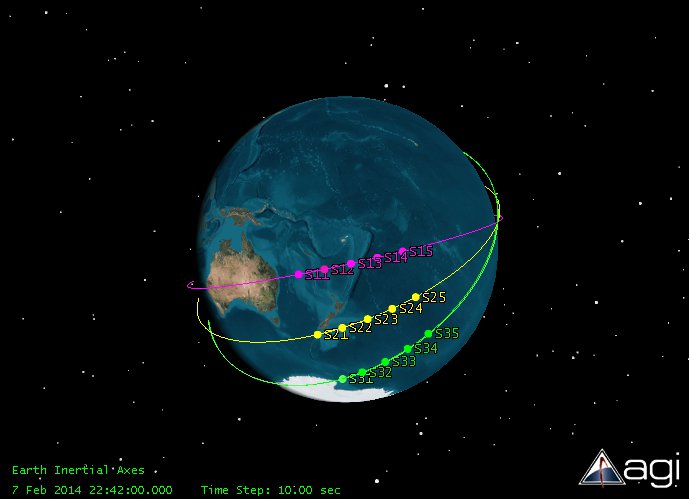}
\caption{Cluster formation flying pattern}
\label{fig:cluster}
\end{figure} 
\item Constellation - This type of formation normally consists of a set of satellites organized in different orbital planes that cover the entire Earth as shown in Figure~\ref{fig:constellation}. Each orbital plane usually contains the required number of satellites in order to provide full coverage for the service being provided.
\begin{figure}[h]
\centering
\renewcommand{\figurename}{Figure}
\includegraphics[width = 3in, height = 2.5in]{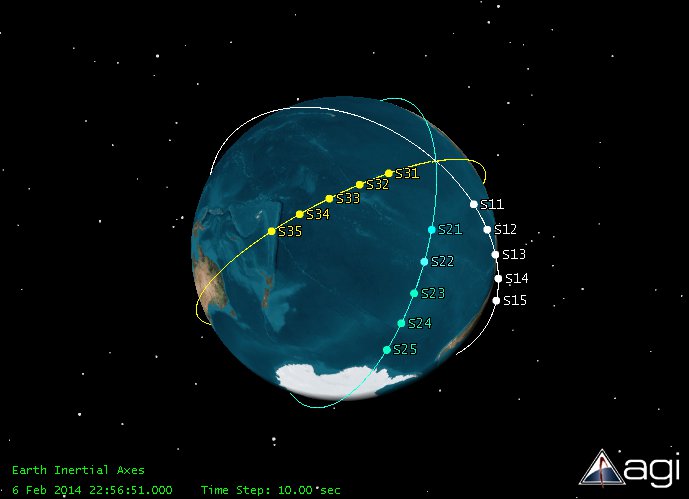}
\caption{Constellation formation flying pattern}
\label{fig:constellation}
\end{figure} 
\end{enumerate}

These multi-satellite configurations fall into a general class defined as Distributed Space Systems (DSS). Fractionated spacecraft and satellite swarms are the new cutting edge technologies for future space missions, which are also subsets of the DSS. A satellite swarm is defined as a set of agents which are identical and self organizing that communicate directly or indirectly and achieve a mission objective by their collective behavior~\cite{formationpatterns}. Fractionated spacecraft is a new satellite architectural model where the functionalities of a single large satellite are distributed across multiple modules, which interact using wireless links~\cite{formationpatterns}. Unlike other multi-satellite configurations, the different modules of this type of configuration are highly heterogeneous corresponding to the various subsystem elements of a conventional large satellite.

\subsection{Inter-Satellite Communications} 
Extending networking to space requires inter-satellite communications which will enable autonomous transfer of data and hence being analogous to terrestrial Internet with autonomous transfer of data with minimum human intervention.  
Inter-Satellite Communications (ISC) assist in performing advanced functions including, for example, distributed processing, servicing or proximity operations, autonomous applications, and fractionated operations as described in Section 1.3. It facilitates in eliminating the use of extensive ground based relay systems and worldwide tracking systems. It also helps to provide attitude control and maintain the relative distance between small satellites. Inter-satellite communications support transmission with high capacity and data rates, real time data delivery, and also can provide absolute interoperability among various spacecraft within the system. The ISC enables navigation and formation control by exchanging the attitude and position information and also maintains time synchronization between the spacecraft. Consequently, inter-satellite communications enable multiple satellite missions for Earth observations and inter-planetary explorations and observations~\cite{satellitesensormotes}.\\

The current state of the art for small satellite communications is a one hop link between satellite and ground stations. Space agencies have developed future missions involving multiple satellites with inter-satellite communications intended to achieve mission objectives: for example, gravity mapping, servicing or proximity operations, etc. Examples of multiple satellite missions with inter-satellite communications are Iridium, Orblink, Teledesic~\cite{Kusza_intersatellitelinks:}, Proba-3~\cite{Proba3reference}, Edison Demonstration of Smallsat Networks (EDSN) mission~\cite{EDSNreference}, ESPACENET~\cite{espacenet_ref}, NASA's Autonomous Nano-Technology Swarm (ANTS)~\cite{ants_ref}, and QB-50 mission~\cite{QB-50reference}. However, much work remains in-order to achieve an in-depth understanding of the communication architecture in an absolutely autonomous and heterogeneous network of small satellites.\\

To facilitate ISC between small satellites, we propose to use the OSI model as a framework to serve as a reference tool for communication between different devices connected in a network. This divides the communication process into different layers. It is a conceptual framework that helps to understand complex interactions within a network. The OSI model has seven layers: physical, data link, network, transport, session, presentation, and application~\cite{tananbaum}. Each layer has well defined functions and offer services to the layers above and below it. It can be used as a framework for the network process for inter-satellite communication in small satellite systems. The small satellite system typically consists of multiple mobile nodes forming a dynamic network topology.  However, these systems have limitations both at the transmitting and receiving end: for example, limited power, mass, antenna size, on-board resources, computing capabilities, intermittent communication links, etc. The overall architectural design of the various layers of the OSI model largely determines the performance of the entire system taking into account the various system constraints. This will enable the expansion of inter-networking to deep space with lower operational costs. The advancements in communication and navigation technology will allow future missions with enhanced capabilities that will enable high bandwidth communication links.  \\

The next section and following subsections will explain the design parameters pertinent to the different layers of the OSI model. The focus of this paper is on the first three layers of the OSI framework since the design criteria for the upper layers is mission/application specific whereas  for the lower layers, it can be generally characterized. Qualitative design approaches taken by various research groups in the area of inter-satellite communications for small satellite systems are also discussed.
\section{Design of Various Layers of the OSI Model}    

\begin{figure}[h]
 \centering
 \renewcommand{\figurename}{Figure}
 \includegraphics[width =3.2in]{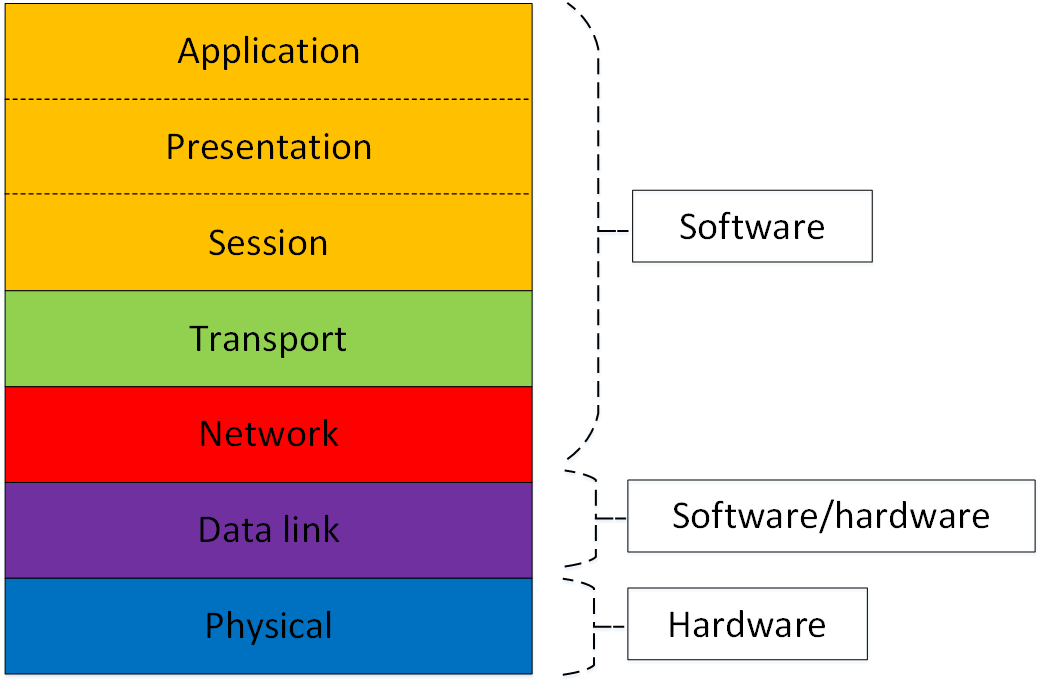}
 \caption{Framework for inter-satellite communication~\cite{tanyadesignissues}}
 \label{fig:osi}
 \end{figure}

\begin{figure*}
 \centering
 \renewcommand{\figurename}{Figure}
 \includegraphics[width =5.8in, height = 3in]{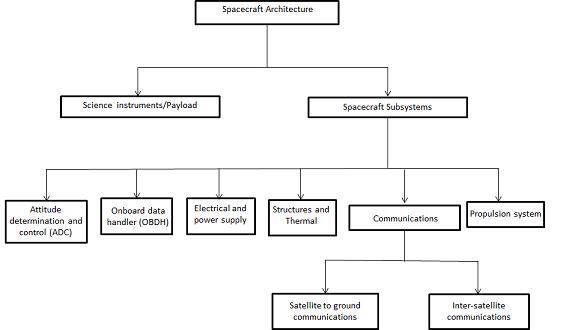}
 \caption{Overall small satellite system architecture}
 \label{fig:subsystems of osi} 
  \end{figure*}

 The overall architecture for inter-satellite communication can be developed using the Open System Interconnection (OSI) model or its derivatives~\cite{sysconconf, useofOSI}. For small satellite systems, the upper three layer functionalities of the OSI model can be merged as shown in Figure~\ref{fig:osi}, which can be implemented using software programs~\cite{tanyadesignissues}. Figure~\ref{fig:subsystems of osi} shows the different subsystems involved in the overall architecture of a spacecraft. 

\subsection{OSI Physical Layer} 
The physical layer is the lowest layer in the seven-layer OSI model for inter-networking of devices which consists of the basic hardware technologies for transmission across a network. The physical layer defines the means of transmitting raw data bits rather than logical data packets over a hardware transmission medium. Specified in this layer are various low-level parameters such as electrical connectors, transmission media, modulation schemes, transmission frequency, specification of signal strength, bandwidth, etc. A more detailed description of various parameters used in small satellites are described below. In this section, we have also reviewed different antenna types used in small satellites. 
\subsubsection{Frequency allocation and data rate}

The regulatory community has allocated a wide range of frequencies for services that support ISC for various applications. It is not necessary to pursue new frequency allocations since existing spectrum should be sufficient to meet expected demands till 2020~\cite{architecturereport01}. The required bandwidth depends on several factors: mission operational requirements and objectives, the type and amount of data transmitted using inter-satellite links, frequency of data transmissions, inter-satellite link statistical parameters (orbital constraints, spacecraft size and power, costs, cross link path lengths), number of simultaneous inter-satellite communications, propagation effects including free space path loss, Radio Frequency (RF) component availability, directionality of the links, etc.

\hskip 0.15in

\begin{table*}[t]

\caption{Frequency bands used for communication~\cite{SMAD}}
\label{tab:frequency bands}

\label{tab1}

\vskip 0.15in

\begin{center}
\centering
\begin{tabular}{| >{\centering\arraybackslash}m{4.5 cm} | >{\centering\arraybackslash}m{4.5 cm} |>{\centering\arraybackslash}m{4.5 cm} |}

\hline

\textbf{Frequency Band} &  \textbf{Uplink Frequency, GHz} & \textbf{Downlink Frequency, GHz}\\

\hline
\textbf{UHF} & \textbf{0.2 - 0.45} & \textbf{0.2 - 0.45}  \\
\hline
L & 1.635 - 1.66 & 1.535 - 1.56 \\
\hline
\textbf{S} & \textbf{2.65 - 2.69} & \textbf{2.5 - 2.54} \\
\hline
C & 5.9 - 6.4 & 3.7 - 4.2  \\
\hline
\textbf{X} & \textbf{7.9 - 8.4} & \textbf{7.25 - 7.75}  \\
\hline
Ku & 14.0 - 14.5 & 12.5 - 12.75  \\
\hline
Ka & 27.5 - 31.0 & 17.7 - 19.7\\
\hline

\end{tabular}

\end{center}

\vskip 0.1in

\end{table*} 
As per Shannon theorem~\cite{futurenanosats1}, channel capacity can be increased by varying channel bandwidth and signal to noise power ratio. The bandwidth can be increased by choosing appropriate modulation and coding schemes. The signal to noise power ratio can be increased in several different ways including increasing antenna gain, increasing the RF output power of the transmitter amplifier, and decreasing the temperature of the system for reducing noise. However, options of increasing antenna gain are limited because of the size constraints on small satellites. It is concluded in~\cite{futurenanosats1} that increasing the bandwidth is a better option than the signal to noise ratio for small satellites provided the extra bandwidth is available. It has also been shown in~\cite{futurenanosats1} that higher data rates can be achieved by transmitting the data as bursts rather than as continuous downlink.   

The total maximum data rate that can be supported in various configurations can be derived using the mathematical equation given in~\cite{architecturereport01}. The total maximum data rate in turn determines the bandwidth requirements. 
\begin{equation*}
\text{Total Maximum Data Rate (MDR)} = \sum_{i=1}^{3} N_j * [\sum_{i=1}^{3} P_{ij} * MDR_i]
\end{equation*}

\noindent where the subscript \emph{i} corresponds to the bandwidth requirements; \emph{i} = 1 for narrow ($<$ 100 Kbps), \emph{i} = 2 for medium (100 Kbps to 10 Mbps) and \emph{i} = 3 for wide ($>$ 10 Mbps) bandwidth ranges respectively. Subscript \emph{j} corresponds to the network architecture; \emph{j} = 1 for constellation, \emph{j} = 2 for centralized formation and \emph{j} = 3 for distributed formation respectively. $P_{ij}$ corresponds to the probability that a particular maximum bandwidth is required for a mission, and $N_{j}$ denotes the number of simultaneous cross link communications possible in the system. The $MDR_i$ is the maximum data rate for each bandwidth category (science, health and status, navigation, and command data), which is explained in detail in~\cite{architecturereport01}. In~\cite{rnew5}, authors propose a rate control protocol for interplanetary networks which are characterized by extremely long propagation delays, high link error rates, asymmetrical bandwidth, and outages. The paper proposed a novel rate control protocol called RCP-Planet to overcome these challenges utilizing a novel rate probe mechanism and rate control schemes that adapt to the available bandwidth. They also proposed to use Tornado codes for packet-level Forward Error Correction because of their fast encoding and decoding speed. \\ 

The inter-satellite links are subject to interference which is a function of the number of transmitters operating in the same frequency band, spatial distribution of the satellites, antenna design, and operational time periods. However, number of cross links that can be simultaneously operated can be increased significantly using appropriate multiple access techniques and type of antennas (directional antennas). The probable number of simultaneously operational inter-satellite links for each mission is estimated based on the number of spacecraft, architecture, and objectives considering the available types of multiple access alternatives and the cost associated with these techniques~\cite{architecturereport01}. \\

The majority of cubesat programs utilize the Ultra High Frequency/Very High Frequency (UHF/VHF) transceivers for downlink communication with no inter-satellite links~\cite{survey01}. Frequencies ranging from VHF (30 MHz) to Ka band (40 GHz) are feasible for inter-satellite communications provided the cubesat has enough power available to support this high frequency transmission and reception. Increasing the frequency for inter-satellite communications reduces the size and mass of the transceivers, and also scales down the antenna size. This helps in achieving high bandwidth which is suitable for applications that require high data rates. The frequency bands bolded in Table~\ref{tab:frequency bands} are used for small satellite communications currently.\\

\subsubsection{Modulation and coding schemes}

Binary Phase Shift Keying (BPSK) is presently the preferable choice for small satellites because these coherent systems require the least amount of power to support a given throughput and bit error rate. Non-coherent systems, for example, Frequency Shift Keying (FSK) requires higher transmitter power compared to BPSK to support the same throughput even though it provides instantaneous communication. However, BPSK systems have inherent delays due to the time it takes to coherently lock to the incoming signal at the receiver side~\cite{survey02}. Quadrature Phase Shift Keying (QPSK) or offset-QPSK modulation techniques are also preferred, as it is a more bandwidth efficient type of modulation than BPSK, potentially twice as efficient. Though, at the receiver end, phase distortion caused due to channel can degrade the performance which can be overcome by differential PSK. Higher order PSK techniques enhance the spectral efficiency, however, the symbols are very close together which can be easily subjected to noise and distortion. Such a signal has to be transmitted with extra power to spread the symbols compared to the simpler schemes like BPSK or QPSK schemes. Therefore, there is a trade off between the spectral efficiency and power requirements. \\

Forward error correction coding significantly reduces the signal to noise ratio requirement, thereby reducing the required transmitter power and antenna size. The coding schemes involve adding parity bits into the data stream at the transmitter. At the receiver end, the parity bits enable the receiver to detect and correct for a limited number of bit errors caused by noise or interference in the channel. A common type of error correction coding scheme used is convolution coding with Viterbi decoding. A 1/2 rate convolution code is implemented by generating two bits for each data bit and hence the data rate is half the transmission rate. The receiver demodulates and stores the data. It is then compared with the coded sequences which could have been transmitted~\cite{SMAD}. Another coding scheme used for deep space network and satellite communications is the Low Density Parity Check (LDPC) code. The main advantage of the LDPC code is that it provides a good performance close to the Shannon capacity for varying noise levels. In~\cite{ldpccodes}, the authors have proposed (512, 256) LDPC code which significantly improved the bit error rate of the system with three micro satellites flying in formation in the same orbit. The performance of various modulation and coding schemes are explained in detail in~\cite{SMAD} \\

\subsubsection{Link design}
Link design analysis relates the transmit to the receive power and shows in detail the feasibility of a given system. A link budget calculation provides excellent means to understand the various system parameters, where by a trade off between the desired system performance at a given cost and level of reliability of the communications link can be obtained~\cite{link_design}. The Space Mission Analysis and Design (SMAD) provides a detailed illustration for link design analysis~\cite{SMAD} and is reiterated below: 
\begin{enumerate}[label=(\alph*)]
  \item Identifying communication requirements - This step involves developing mission requirements including the number of satellites, orbital parameters, mission objectives, etc, and also involves identifying the location of ground stations and relay stations. 
  \item Determining data rates for inter-satellite links as well as uplink/downlink - It is required to determine the data rates, sampling rates, quantization levels, and the number of bits per symbol. This in turn depends on the mission objectives, the type of data exchange between the satellites, the frequency of data transmission, and the available bandwidth.
  \item Design of each link - Each link (cross links and uplink/downlink) can be designed depending on numerous parameters: for example, frequency band of transmission, the modulation and coding techniques used, antenna size, gain, beam width constraints and interference effects, estimation of atmospheric or rain absorption, transmitter power and received noise.
  \item Size of the communication payload subsystem - The size of the communication system depends on the payload antenna configuration, the size and mass of the antennas, transmitter mass and power, payload mass and power, and power required for antenna transmission and reception.\\ 
 \end{enumerate} 
 
\subsubsection{Antenna design in small satellites for inter-satellite links}

This section gives an overview of the antenna technologies for small satellite applications and is followed by a description of the challenges and constraints of antenna design for small satellites. Various antenna types for small satellite applications are also illustrated. In the literature, multiple antenna techniques for satellite systems have been investigated in~\cite{mimo1, mimo2} and emphasis is given on the viability of Multiple Input Multiple Output (MIMO) antennas on satellites with potential enhancements in terms of channel capacity and link reliability that can be achieved through spatial and/or polarization diversity. However, noting the limited size of small satellites, MIMO antennas might not be the best option for inter-satellite communication in small satellites due to the characteristics of the propagation channel between the satellites. \\
 
There are two antenna techniques than can be used for inter-satellite communications: broad beam width isolated antennas and antenna arrays. The first antenna type to be discussed is broad beam width isolated antennas that provides a more compact and simple architecture, while the antenna arrays offer some advantages in terms of beam steering capability and antenna gain.\\

The first antenna for inter-satellite communications in cubesat platforms is described in~\cite{antennadesign01}. The proposed antenna is a retro-directive array of circularly polarized patches, which has the capability to self-steer a transmitting signal without a prior knowledge of its position. In order to build a feasible antenna for small satellite platform, the authors make use of a heterodyne technique with a phase conjugating mixer: the incoming RF signal and its phase in each antenna element is mixed with a local oscillator at half the RF frequency. This process generates an Intermediate Frequency (IF) signal with frequency similar to the incoming signal, but with a conjugate phase. As the phase gradients of incoming and outgoing signals are opposite, the outgoing wave is steered towards the direction of the source. Proposed RF frequency is 10.5 GHz, with a local oscillator of 21 GHz, so that maximum achievable range is limited by propagation losses.\\

A recent experiment for inter-satellite communications is GAMANET, intended to create a large ad-hoc network in space using ground stations and satellites as nodes with inter-satellite links using S band frequency~\cite{antennadesign02}. The space segment of GAMANET specifies 3 and 6 antennas for 3-axis stabilized and spinning satellites, respectively, for inter-satellite links. The architecture provides capabilities to control multiple antennas in the satellite faces depending on the satellite-to-satellite and satellite-to-ground vectors. According to link budget, a maximum distance of 1000 km between satellites can be achieved using a 3 W transmit power.\\

In~\cite{antennadesign02}, an antenna system with one individual antenna per face of the cubesat is proposed in order to have complete coverage with operational frequency of 2.45 GHz (S band). Individual 5 dBi gain patches are considered and antenna system control is implemented using a beam forming approach. The signals received in each antenna are weighted by a complex factor before combination of the signals. A maximum of three antennas are considered in the combination. Simulation results show that beam forming control antennas present better performance as compared to antenna selection.\\

The antenna design for inter-satellite links of the Orbiting Low Frequency Antennas for Radio Astronomy (OLFAR) mission has gained considerable interest. The OLFAR mission is an initiative to perform ultra-long-wavelength radio using a radio telescope consisting in an aperture synthesis interferometric array implemented with a swarm of nano-satellites, in which each satellite carries one element of the array. Each satellite is a 3U cubesat with dimensions of $10\times{10}\times{30}$ cm. Due to the free-drifting of the satellites, distances and orientation of the satellites varies with time and hence, it is difficult to maintain inter-satellite links in any direction. In the literature,~\cite{antennadesign04,antennadesign05}, antenna designs proposed for inter-satellite links for nano-satellites and cubesats are mostly S-band single-patch antennas and has a trade off between path losses and antenna size. In these cases, antennas are limited in gain, and thus the maximum range and feasible inter-satellite distance is significantly reduced.\\

\emph{Constraints and requirements} - The antenna specifications have to be defined at the earliest stage of the project considering high-level mission requirements. The design and manufacturing of an antenna for inter-satellite communications is critical and must ensure that it is applicable for formation-flying, and distributed satellite missions formed by cubesats and picosatellites.\\
Antenna specifications are imposed by communications, platform, and/or mission aspects.
\paragraph{Mission constraints}
Specifications and constraints imposed by the mission requirements are explained below.
\begin{enumerate}[label=(\alph*)]
\item Angular exploration margin - The antenna beam must be steered within a cone with a semi-angle of $40\deg$ relative to the broad side direction.
\item Knowledge of satellite constellation status - The antenna intend to have beam steering capabilities if the relative positions of the spacecraft in formation is not known. 
\item Space environment - The materials used for antenna must satisfy with the mechanical and thermal constraints of space missions also should be capable of surviving in the radiation environment of the selected orbit.
\item Cost - Low-cost materials and machining procedures can be used for manufacturing the antennas as imposed by the reduced budget of cubesat missions.
\end{enumerate}
\paragraph{Imposed by the platform}
Specifications and constraints imposed by the selected satellite platform are described below.
\begin{enumerate}[label=(\alph*)]
\item Mass and deployer constraints - As small satellites are lighter, the antenna must be made of light materials and must be thin and planar. Materials with high dielectric permittivity will permit a reduction in the antenna size at the expense of higher losses.
\item Aperture size - As small spacecraft area is limited, the antenna or antennas shall fit in the area of a spacecraft side, for example, the antenna shall fit in the 10 cm square size of a cubesat.
\item Power - The antenna can be located in one or several of the square faces of the spacecraft. A trade off between antenna aperture and solar-panel area had to be carried out by the space-systems engineer, taking into account the particular mission and payload requirements.
\item Modularity - A modular antenna with an aperture size that can be configured by adding more modules is an interesting option to fulfill requirements as inter-satellite distance or transfer rate varies.
\item Deployment - The antenna can be attached to the external surface of the spacecraft body so that no deployment mechanism for the antenna would be required to avoid any failure risk.
\item Attitude control accuracy - Due to limitations in the accuracy of the attitude and on-board control subsystem in small satellite missions, the scanning features of the antenna under design must be large, and polarization must be independent of the spacecraft attitude. 
\item Compactness: The antenna must be compact, without moving parts and minimum harness, in order to resist the harsh environment and vibrations during launch.
\end{enumerate}
\paragraph{Imposed by communications}
Finally, there are subsystem specifications that must be considered prior to the preliminary and detailed design of the antenna subsystem. Some of the next requirements and constraints are obtained from a link budget analysis considering the mission requirements.\\
\begin{enumerate}[label=(\alph*)]
\item Frequency band - An Industrial, Scientific and Medical (ISM) radio band can be used by selecting a high frequency band in order to build an antenna satisfying size requirements. However, the selection of the frequency band is also influenced by the availability of RF-COTS (Commercial Off-The-Shelf) components. We must also take into consideration the trade-off between antenna gain and propagation losses.
\item Low loss - Materials and substrates with low dissipation factor (\emph{tan} $\delta$) must be used to avoid degradation of radiation efficiency.
\item Range - According to typical spacecraft separation in formation-flying missions, the maximum range between spacecraft has to be a few kilometers, determined by the orbital characteristics of the mission, which can be different  for various constellation configurations \item Antenna gain - The antenna must facilitate communication between spacecraft for the specified inter-satellite distances. These rates range from 10 kbit/s for single-point Global Positioning System (GPS) processing~\cite{antennadesign04} up to 48 kbytes/s for a relative navigation subsystem using a high-update-rate multi - Global Navigation Satellite System(GNSS) receiver~\cite{antennadesign08}. The minimum bandwidth of the inter-satellite link is generally 1 MHz.
\item Duplex method - Typically, inter-satellite communications require transmission and reception capabilities and can be carried out in the same frequency band in order to have a single antenna for inter-satellite links. For Frequency Division Duplex (FDD) system, transmission and reception bands can be separated using diplexor or circulator. Size and mass of both devices depend on the frequency band which limits the application to small satellite missions. Moreover, a circulator formed by magnetic materials might affect the behavior of attitude control systems based on magnetic torques. In Time Division Duplex (TDD) architectures, transmission and reception paths are separated using an RF switch controlled by the timing signals of the communication systems.
\end{enumerate}
\paragraph{Antenna concepts for inter-satellite links}
As stated above, the antenna concept is derived from mission, platform, and communication requirements. In particular, the antenna concept is influenced by the angular visibility, that is, the angular region where the antenna has to concentrate the radiation. 
It is important to take a closer look at mission architectures with interest in inter-satellite links in order to show the relation between inter-satellite range and control accuracy requirements.\\

\begin{figure}[h]
 \centering
 \renewcommand{\figurename}{Figure}
 \includegraphics[width =3.5in]{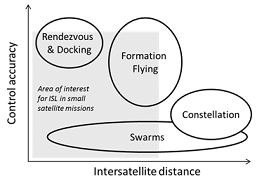}
 \caption{Inter-satellite distance vs control accuracy requirements}
 \label{fig:antennadesign1}
 \end{figure}

Inter-satellite distances and control accuracy requirements can be very different between missions, as shown in Figure~\ref{fig:antennadesign1} (adapted from~\cite{antennadesign09}). In the case of formation-flying missions (e.g., PROBA-3 mission), a reduced number of satellites are concentrated within a small area while maintaining a particular relative position. Constellations used to provide global coverage (e.g., Galileo, GPS, Iridium) are formed by satellites in different orbital planes with inter-satellite distances of several hundreds and even thousands of kilometers. In contrast, satellite swarms are formed by a large number of independent but similar satellites working to achieve a common mission objective with very different inter-satellite distances (e.g., QB-50).\\

From the small satellite's perspective, inter-satellite links are feasible when inter-satellite distances are small, as these platforms are limited by the amount of electrical power they can produce. On the other hand, inter-satellite links are limited by the capabilities of the platform to achieve high control accuracy. The shadow area in Figure~\ref{fig:antennadesign1} represents the potential area to include inter-satellite links in small satellite missions.\\

\begin{figure}[h]
 \centering
 \renewcommand{\figurename}{Figure}
 \begin{subfigure}[t]{0.48\textwidth}
 \includegraphics[width = \textwidth]{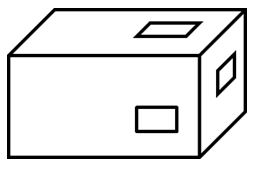}
 \caption{Individual antennas for global coverage}
 \label{fig:antennadesign2}
\end{subfigure}
\begin{subfigure}[t]{0.46\textwidth}
 \includegraphics[width = \textwidth]{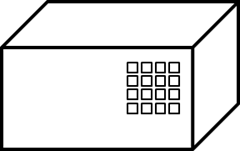}
 \caption{Antenna array with beam steering capabilities}
 \label{fig:antennadesign3}
\end{subfigure}
\end{figure}

From the discussion above, two antenna concepts for inter-satellite links in small satellite missions can be proposed. Figure~\ref{fig:antennadesign2} shows the concept of having individual antennas in orthogonal faces of the spacecraft. For this antenna concept, either antenna selection or beam forming can be implemented, as described in~\cite{antennadesign03}. Figure~\ref{fig:antennadesign3} shows an antenna array located in one of the faces. The antenna array synthesizes a narrow beam with higher gain than individual antennas. The array must have beam steering capability to explore as much angular area as possible. Both concepts could be combined allocating an individual antenna array in three faces, but the available area for solar cells would be limited unless solar cell deployable are used.\\

Table~\ref{tab:antennadesign} compares the features of the two antenna concepts, whereby the features are defined as:.
\begin{table*}[t]
\caption{Comparison of antenna concepts for inter-satellite links for small satellite missions}
\renewcommand{\arraystretch}{1.8}
\label{tab:antennadesign}
\begin{center}

\begin{tabular}{|>{\centering\arraybackslash}m{2.8cm}|m{2.9cm}|m{2.9cm}|>{\centering\arraybackslash}m{5cm}|}
\hline
\textbf{Features} & \multicolumn{2}{|c|}{\textbf{Individual antennas}} & \textbf{Antenna array}\\
\hline
 & \textbf{w/o beam forming} & \textbf{with beam forming} &  \\ \hline
Directivity & Low & Medium & High (depending on antenna array aperture)\\  \hline
Beam steering & Not required & Required & Required \\  \hline
Angular coverage & \multicolumn{2}{|c|}{High} & Medium/low \\  \hline
Occupied area & \multicolumn{2}{|c|}{Medium} & Large , but only in a single face \\  \hline
Inter-satellite range & Low & Medium & Large \\  \hline
Complexity & Low & High & High \\  \hline
Mission & \multicolumn{2}{|c|}{\begin{minipage}{7.5cm}Swarms with low inter-satellite distances and relaxed control accuracy requirements\end{minipage}} & Formation flying missions, Swarms with medium/high inter-satellite distances \\ \hline
\end{tabular}

\end{center}

\end{table*}
\begin{enumerate}[label=(\alph*)]
\item Directivity - Given the reduced area of small satellite sides, the effective aperture and the achievable gain of the antenna is limited if a single antenna is used. Larger apertures and directivity can be obtained with planar antenna arrays allocated in a single face. The use of tri-dimensional arrays allocating individual antennas in orthogonal faces provides lower gain than planar arrays.
\item Beam steering - In case the antenna pattern is directional as in the case of antenna array, the control system of the antenna control unit must steer the beam in the appropriate direction.
\item Angular coverage - The use of planar antenna arrays limits the angular coverage, as it is limited by the radiation pattern of antenna elements. For the standard case of using micro-strip patches as elements of the planar array, the angular coverage is limited to $\pm$40 degrees around the broadside (normal) direction.
\item Occupied area - It compares the area covered by the antenna arrays to the total area of the spacecraft face that can be used for solar panels.
\item Inter-satellite range - Longer inter-satellite link can be established using large antenna aperture The larger the antenna aperture, the longer the inter-satellite link that can be established for the same communication parameters (e.g., bit error rate, bandwidth, signal to noise ratio).
\item Complexity - Antenna arrays with beam forming require the computation of complex weights under different optimization criteria~\cite{antennadesign10}. Thus, a processing unit must be incorporated as part of the antenna subsystem to extract information of the inter-satellite link direction and the calculation of complex weights. The required hardware depends on the beam forming algorithm and computational load increases with the number of antennas in the array~\cite{antennadesign11}.
\item Mission - Each antenna concept is more adequate for a space segment architecture. Low-directivity antennas are good candidates for missions with relaxed control accuracy requirements and low inter-satellite distances. On the other hand, arrays can also be used in formation flying missions with stringent control accuracy.
\end{enumerate}

\subsection{OSI Data Link Layer}

The data link layer is one of the most complicated layers of the OSI model due to complex functionalities in a network with multiple satellites sharing the same medium. This layer is responsible for various functions such as framing, physical addressing (Medium Access Control/MAC address), synchronization, error control, flow control, and multiple access. It is divided into two sub layers: logical link control layer (deals with flow and error control) and medium access control layer (deals with actual control of media). The multiple access protocol design plays a vital role in the performance of the entire system. The basic function of a MAC protocol is to avoid collision by arbitrating the access of the shared medium among the nodes in the network~\cite{Zengstext}.\\

A typical scenario in a wireless sensor network consists of a large number of nodes that need to communicate using a single channel. Generally, transmission from any node can be received by all other nodes in the network. Therefore, if more than one node in the network attempts to transmit at the same time, collision occurs, which will result in the loss of data packets. The receiving node cannot interpret the data which is being transmitted and such a situation is called collision~\cite{Zengstext}. In order to avoid collisions, the nodes in a network should follow some set of rules or protocols that would allow fairness among the nodes for accessing the channel, and also will result in the effective channel utilization. The protocols determine which node in the network gets access to the shared channel at a given time and for a given duration, thus avoiding collision. A large number of satellites can be deployed as a satellite sensor network, which applies the concept of terrestrial wireless sensor networks to LEO spacecraft for various space missions~\cite{tanyaWSN}.\\

In order to improve the performance of the network, the MAC protocols should be designed taking into account mission specifications such as, mission application, network topology, number of satellites, etc. Also, it is important to consider several system constraints of small satellites, for example, limited on-board power and computing resources. Depending upon the numerous mission applications, the MAC protocols are required to autonomously adapt to several factors like scalability, adaptability, channel utilization, latency, throughput, and fairness~\cite{jamalipour}.These factors are explained in detail below.

\begin{enumerate}[label=(\alph*)]
\item Energy efficiency - The energy consumed per unit of successful transmission is defined as energy efficiency. The nodes in a wireless sensor network are typically battery powered and often placed in remote locations where human intervention is not possible. Therefore, it is important to use the battery power effectively. The MAC protocol should be designed in such a way to ensure lower energy usage in the nodes, and thereby of the entire network. 
\item Scalability and adaptability - Scalability is defined as the ability of the network to adapt to the changes in the size of the network. There may be many applications where a set of satellites may join an already established network. The MAC protocol should be able to adapt to such changes in the network size. Adaptability refers to the capacity to accommodate changes in the node density and overall topology of the network. In any network, satellites can join, fail, or reconfigure themselves into different topologies to which the MAC protocol should adapt efficiently.
\item Channel utilization - It refers to the effective bandwidth utilization. The MAC protocol should be designed such that the bandwidth, which is limited, is utilized in an efficient manner. 
\item Latency - The length of time it takes for a data packet to reach its destination successfully is defined as latency. The importance of latency depends upon the mission type. For real time applications where we need continuous transfer of data, latency should be minimal. Hence, the MAC protocol design should consider the different types of missions.
\item Throughput - The amount of data successfully transmitted across the channel in a given time and usually expressed as bytes/second. It depends on numerous factors like latency, communication overhead, channel utilization, etc~\cite{Zengstext}. 
\item Fairness - The MAC protocol has to be designed in a manner such that it ensures equal opportunity for all satellites in a network to get access to the channel. It is important not only to guarantee per-node fairness, but also to ensure the quality of service of the entire system which is defined as the efficiency and it is in trade-off with fairness.
\end{enumerate}

There are two different types of multiple access protocols for handling collision of data packets: contention based and conflict-free protocols. According to contention based protocol, satellites compete for the channel, and when collision occurs the protocol carries out a collision resolution protocol. Numerous contention based protocols have been proposed in the literature~\cite{wirelesscomm}, for example, ALOHA, CSMA (Carrier Sense Multiple Access), BTMA (Busy Tone Multiple Access), ISMA (Idle Signal Multiple Access), etc. The collision free protocols ensure that collision of data packet never occurs. Some of the basic protocols of this type are TDMA (Time Division Multiple Access), FDMA (Frequency Division Multiple Access), and CDMA (Code Division Multiple Access). The OFDM (Orthogonal Frequency Division Multiple Access) and SDMA (Space Division Multiple Access) are the other two variations which have been introduced recently.\\

Research is being conducted on various multiple access methods for inter-satellite communications in small satellite systems. The authors in~\cite{ISC05} propose IEEE 802.11 physical and MAC layers for space based Wireless Local Area Network (WLAN) by optimizing the four main types of inter-frame spacings defined in IEEE 802.11: the Short Inter-Frame Space (SIFS), the Distributed Co-ordination Function (DCF) Inter-Frame Space (DIFS), the Point Co-ordination Function (PCF) inter-frame space, and the Extended Inter-Frame Space (EIFS). They investigated the impact on IEEE 802.11 standards for LEO satellites deployed in polar and inclined orbits for two different formation flying scenarios: the triangular and the circular flower constellations. Both formations are modified to include a master-slave configuration, with the master satellite acting as the access point (similar to terrestrial WLAN) and slave satellites act as mobile nodes. They analyzed the various configurations using extensive simulations in ATK Systems Tool Kit (STK)~\cite{stk}. A Doppler shift of 100 KHz and 50 KHz were experienced for the triangular and the circular formations respectively, which is within the specifications of the IEEE 802.11 for mobility. They also analyzed different scenarios by placing both formation flying patterns in frozen orbits (orbits maintaining almost constant altitudes over any particular point on the Earth's surface). The results indicate that the Doppler shift is considerably less in frozen orbits compared to the sun synchronous orbits. In conclusion, the slave satellites are locked tightly to the master satellite when placed in frozen orbits in comparison to sun synchronous orbits. The IEEE 802.11 standards are designed for terrestrial applications for outdoor distances of 300 meters. In LEO networks, the inter-satellite distance can range from ten to several thousands of kilometers and propagation delays are in the order of milliseconds which is much larger than the delays present in terrestrial mobile networks. Hence, completely re-defined IEEE 802.11 MAC timings for inter-satellite ranges are proposed. Depending on the maximum distance between the satellites, propagation delay increases and all the parameters of IEEE 802.11 standards have to be re-defined. The OPNET~\cite{opnet} simulation results shows that the DCF MAC suffers from degradation at large distance. However, constant throughput is achieved if optimum inter-frame space values are used. They also suggested that by finding optimum probability ratio between the collided packets and the successful packets, and relating the success ratio to the optimal contention window (CW), the satellites can adjust their CW minimum values adaptively thereby operating at optimal conditions. Integration of IEEE 802.11 MAC with smart antennas (Adaptive Antenna System, AAS) is also proposed with an increase in performance gain as compared to omni-directional antennas. The proposed antenna scheme ensures high spectral efficiency by means of increased protection against fading, thermal noise, and multiple access interference. It is concluded that the IEEE 802.11 can be extended to longer inter-satellite link with minimum degradation of throughput.\\

A MAC protocol based on Carrier Sense Multiple Access with Request-to-Send and Clear-to-Send protocol (CSMA/CA/RTS/CTS) is proposed in~\cite{ISC02, ISC08} for various formation flying patterns of small satellites. At the physical layer, depending on the type of formation flying pattern, smart antennas are suggested. The CSMA/CA/RTS/CTS is traditionally designed in such a way that the RTS, CTS control commands and data are transmitted from a source to a destination in an omni-directional way. However, depending on the network topology of small satellites in various configurations, smart antennas can be used to transmit RTS, CTS and data, thereby saving power which is a premium component for small satellites. For example, for leader-follower system all satellites are deployed in a single orbit, separated from each other at a specific distance, hence RTS, CTS control commands and data can be sent using directional antennas. Extensive simulations are executed and it is concluded that the proposed MAC protocol is suitable for missions that do not require tight communication links. The proposed protocol is discussed in detail in Section 4.3.1.     \\

The communication mechanisms for a constellation of 50 cube satellites, the QB-50 project, are investigated in~\cite{ISC01}. They evaluated Transmission Control Protocol (TCP) and User Datagram Protocol (UDP) for a network of satellites. Two different network topologies are evaluated: the first scenario is a ring of 50 equally spaced cube satellites, and the second is a 10,000 km cubesat string. The proposed satellite constellation analysis is based on the following facts: The constellation is placed in one polar orbital plane at an inclination of 79 degrees, at an altitude of 300 km. There were nine ground stations, most of them are from the Global Educational Network for Satellite Operations (GENSO, \cite{genso1, genso2}) project. The uplink and downlink data rates are 9600 bps and inter-satellite link data rates will take different values in the simulations; 0.5 kbps, 1 kbps, 3 kbps, 6 kbps, 8 kbps and 10 kbps. The traffic is assumed to be Constant Bit Rate (CBR) with packets of size 210 bytes, transmitted every second. The queue has a capacity of 50 data packets following tail drop policy. The Tool Command Language (TCL) based Network Simulator (NS-2) and satellite visualization software are used for simulating the proposed environments~\cite{ISC01}. The authors analyzed the system using three different parameters: throughput arriving to ground station located at Lima, Peru; delay of the packets for
the flow from Satellite 1 until they are received in the ground station; packet loss rate due to channel errors or congestion only. It has been demonstrated that TCP maintains optimum throughput throughout the simulation time and has less packet loss unlike UDP. However, there is a greater delay and lesser packet loss associated with TCP compared to UDP. Also, it is proposed that the traffic distribution was better for the network topology proposed in the first scenario because of the great symmetry level compared to the second scenario. The proposed multiple access protocol is AX.25. Thus, the advantages of existing terrestrial protocols are utilized and tried to implement in space. The data packets should be exchanged in a timely manner to estimate the inter-satellite distance, and thereby terrestrial protocols may not be applicable to space based networks. It is concluded that TCP is an ideal choice for reliable and error free communication where as UDP would be a good choice for quick transmission.    \\

The capabilities of Code Division Multiple Access (CDMA) for Precision Formation Flying (PFF) missions are investigated in~\cite{ISC04, ISCref9}. The PFF missions require high navigational accuracy and high measurement update frequency and hence primary concerns in PFF missions are time criticality and operational flexibility. The main requirement in PFF missions is to acquire and maintain the spacecraft in the relative geometry. The data exchanged between the satellites should be arrived in time to estimate the inter-satellite distance. A relative navigation filter, for example, extended Kalman filter can be used to account for the relative navigational errors of the spacecraft which employs a numerical integration scheme. The measurements used in the filter is given by the spacecraft which can be either the unambiguous
coarse code or the ambiguous precise carrier phases. For PFF missions, relative navigational measurements changes as the formation evolve through different phases of precision formation, requiring different levels of position sensing and control maneuvering. Spacecraft can be considered as free flying entities that aggregate into a desired spatial arrangement thereby eventually discovering other spacecraft which may already be a member of a multi-spacecraft network, hence establishing ``complete connectivity". This condition is defined as formation acquisition, where the system evolves in to a centralized graph with one spacecraft chosen to be the reference for a particular time period and subsequently enabling various science missions, for example, multi-point remote sensing. A half-duplex CDMA is selected as a suitable network architecture since it enables both code and carrier phase measurements, and also supports reconfigurability and scalability within space based sensor networks. The authors also propose to rotate the functionalities of the mother satellite among other satellites within the network (roles rotating architecture). It provides better capabilities compared to fixed time slot TDMA by obtaining measurements from all the spacecraft in a single time slot using CDMA strategy. The signals transmitted from spacecraft need not have to start at the same time thus allowing scalability. Using CDMA, GNSS technology can be utilized to a large extent thereby improving ranging accuracy. The limitations of using CDMA in terms of Multiple Access Interference (MAI) as well as near far problems are also discussed for a lower Earth circular mission with 5 satellites, one mother and four daughter satellites. The effect of Doppler frequency is also analyzed, and it is shown that reducing the energy per bit to noise density ratio will lead to reduction in MAI, but limits the inter-satellite separation diversity and the maximum number of satellites in the network. The MAI, along with Doppler effects and near far problem, worsens navigational accuracy which is a critical issue in precision formation flying missions. The effects of MAI is studied in a NASA's mission, Magnetospheric Multi-Scale (MMS) formation, consisting of four identical satellites in a tetrahedral geometry. It is observed that for the MMS mission, Doppler
offset is beyond the crossover sensitive zone for a substantial time of an orbit period, resulting in smaller cross-correlation errors. When the satellites are in close proximity, an adaptive power control mechanism, lowering the power of the transmitted signals, can be used to minimize the effect of near-far problem. It is suggested that, for PFF missions with tight control periods, the requirements on high update rates of the code and carrier phases need to be carefully considered. \\

A hybrid combination of CSMA and TDMA called Load Division Multiple Access (LDMA) is investigated in~\cite{ISC03}. It combines the advantages of both CSMA and TDMA protocols depending on the level of competition in the network. The proposed protocol operates in two different modes, Low Contention Level (LCL) and High Contention Level (HCL) mode. In case of low network congestion, the network uses CSMA and for high network congestion, TDMA is used, thereby improving the communication performance. For the proposed protocol, unlike the traditional CSMA protocol, each node is assigned a priority. The owner of the slot has high priority to transmit over the non-owner of the slots thereby reducing collisions. The slots can be used by non-owners, if the owner of the slot does not generate data to be transmitted. The LDMA protocol does not utilize the RTS/CTS control commands and hence the network congestion is directly proportional to the conflict probability. The switching between CSMA and TDMA is based on the number of conflict frames received. If the master node receives \emph{N} conflict frames, it broadcasts a notification indicating that the system is in HCL mode. Accordingly, the nodes switch to TDMA mode thereby achieving high channel utilization and throughput without the need of accurate timing synchronization. The system model consists of a large number of satellites in a circular formation with a base radius of 1400 km. The circular formation collapses into a line as the satellites approaches poles. There is a master satellite and the other satellites are around the master satellite at a specific altitude. The master satellite collects data from other satellites and transmit it to the ground station. The authors realized the LDMA protocol using OMNET++ platform and Systems Tool Kit (STK). The system performance is evaluated using three different measures; channel utilization, collision probability and throughput, and the results are compared with pure CSMA and TDMA protocols. Through extensive simulations, it is shown that, LDMA achieves maximum channel utilization of 72\% with increasing traffic in the network, compared to pure CSMA (44\%) and TDMA (61\%) systems. It is observed that LDMA achieves high throuput compared to TDMA and CSMA and has much lower collision probability compared to pure CSMA. However, for a large and scalable network of small satellites, the LDMA protocol may not be a good choice since the performance of CSMA deteriorates with the increase in number of satellites. Also, for TDMA, the master satellite may not be able to cover the whole system within its transmission range because of the low transmission power and time scheduling will be difficult in a scalable network.\\
\begin{table*}[!ht]
\caption{Comparison of different MAC protocols for small satellite systems}
\label{tab:mac protocol comparison}
\renewcommand{\arraystretch}{1.3}
\begin{center}
\begin{tabular}{| >{\centering\arraybackslash}m{2.2cm} | >{\centering\arraybackslash} m{1.6cm} | >{\centering\arraybackslash}m{2.7 cm}|  >{\centering\arraybackslash}m{2.5cm} | m{3cm} | m{3cm} |}

\hline

\textbf{Protocol} & \textbf{Topology} & \textbf{Synchronization} & \textbf{Contention based/Conflict free} & \textbf{Advantages} & \textbf{Disadvantages} \\

\hline

CSMA/CA with RTS/CTS \cite{ISC05, ISC02, ISC08} & Distributed & No & Contention based & Loosens the synchronization requirements & High offered load is challenging, Not suitable for missions requiring tight communication links  \\

\hline
TCP/UDP with AX.25 \cite{ISC01} & Distributed & No & Contention based & Loosens the synchronization requirements & Not well suited for operation over noisy and band limited links  \\
\hline
TDMA & Centralized & Yes & Conflict free & High bandwidth efficiency & Not suitable for a system with large number of satellites  \\
\hline
Half duplex CDMA \cite{ISC04, ISCref9} & Centralized & Yes & Conflict free & Less delay, High throughput & Near far problem and MAI affect the performance, Limits the number of satellites in the system \\
\hline
LDMA (hybrid of CSMA and TDMA) \cite{ISC03} & Distributed and Centralized & Yes & Contention based/Conflict free & LDMA achieves maximum channel utilization compared to pure CSMA and TDMA & May not be a good choice with increase in the number of satellites, time scheduling is difficult in a scalable network \\
\hline
Hybrid of TDMA and FDMA \cite{ISC06} & Centralized & Yes & Conflict free & High band width efficiency, eliminates the hidden and exposed node problem & Not suitable for dense and heavily loaded network \\
\hline
Hybrid of TDMA and CDMA \cite{ISC07} & Centralized & Yes & Conflict free & Less delay, High throughput, Suitable for scalable and reconfigurable small satellite missions & Strict synchronization required \\

\hline
\end{tabular}

\end{center}
\end{table*}

A combination of TDMA and CDMA for a cluster of satellites is proposed in~\cite{ISC07}, consisting of a TDMA-centric and CDMA-centric approach. The dynamic and unpredictable behavior of space environments would lead to delayed and disrupted communication links. Future space projects can be envisioned as different phase missions where satellites may be deployed at discrete time instances to accomplish mission objectives and unexpected failures can occur in the network. Thus, the MAC protocol must be able to handle dynamically changing cluster geometries, which may be unpredictable. Taking into account all these objectives, the authors in~\cite{ISC07} have proposed to divide the whole network into clusters and to implement a master-slave model with each cluster having a master satellite and several slave satellites. In order to prevent single point failures of master satellites, re-clustering of the network is suggested using closeness centrality algorithm. Extensive simulations are performed based on the CDMA-centric frame structure and it is shown that the hybrid TDMA/CDMA protocol has high throughput and delay as compared to other protocols. The hybrid TDMA/CDMA protocol is explained in detail in Section 4.3.2.\\

A hybrid combination of FDMA/TDMA is proposed in~\cite{ISC06} by modifying WiMedia MAC and PHY layer parameters to meet the requirements of inter-satellite networking. The authors proposed two dimensional time-frequency slots for communication between satellites thereby addressing challenges of efficiency and flexibility. In the 2D super frame structure, first few slots are allocated for beacon signals and then the time-frequency slots are allocated to a given node to communicate with other nodes. Two different ranges of operations are defined; Normal Range (NR) in which satellites are expected to operate with in 10 km and Extended Range (ER) where modules are separated by hundreds of kilometers. For ER, high data rates may not be available, however, the satellites are expected to be able to maintain basic command and control communication. They also proposed two different modes of operation Single Mode (SM) and Dual Mode (DM). In SM, there is only one type of super frame for both NR and ER, utilizing all frequency sub-bands. For dual mode, two different frame structures are defined, one for NR and another for ER. A satellite goes in to single mode if all the satellites are in either normal range or extended range. When the system is in Dual mode, a designated node is chosen to monitor the extended range communication by checking on remote nodes and releasing near modules to communicate at higher data rates. For long range links (inter-satellite distances of more than 10 km), sub-bands are added to improve network capacity through maximum utilization of spectrum. Also, for changing mission requirements, it is suggested to dynamically adjust the MAC parameters for various operations i.e., super frame
structure, normal and extended ranges of operations, and dual versus
single mode of operations. Also, it is shown that simplex communication is a less costly approach for inter-satellite networking.\\

To date, the link protocol standards established for space flight communications by the Consultative Committee for Space Data Systems (CCSDS)~\cite{ccsdsnew} have not been widely used in smallsat mission operations.  Many of those standards were primarily designed for use in deep space missions over very long signal propagation delays, an environment that is quite different from low-Earth orbit.  However, missions employing smallsat technology for deep space science and exploration are now being developed~\cite{ccsds1}, and CCSDS link protocols such as Proximity-1~\cite{ccsds2} are in any case suitable for spacecraft in planetary orbit, so the CCSDS standards may play an increasingly significant role in future smallsat communications.\\

The selection of MAC protocols largely rely upon the mission objectives and the number of satellites in the whole system. Table~\ref{tab:mac protocol comparison} shows the various protocols suggested for inter-satellite communications in the literature.

\subsection{OSI Network Layer}

The network layer is responsible for data packet routing. Routing is the process of moving information across an inter-network from source to destination nodes, whereby many intermediate nodes maybe encountered. Routing protocols use metrics to validate the best path for passing information. There are various metrics for determining the optimum path such as delay, bandwidth, path reliability, link status, load on a particular link, hop count, and bandwidth. Two most important network optimization objectives are transmission power and time. When the data packet is transmitted to the destination satellite by multi-hopping, there is a significant reduction in the power for communication. It is important to determine an optimum route for a data packet that is being transmitted between a sender and a receiver. For leader-follower formation flying pattern, Bellman Ford algorithm is proposed~\cite{ISC02}, where the routing metric is the minimum number of hops between the sender and receiver. Proactive and reactive routing schemes can be used depending on the topology of the whole network. The choice of the routing scheme is also dependent upon the mission requirements, whether it is possible to use a completely distributed or centralized system. For proactive scheme, each satellite knows the entire network topology, and whenever a satellite needs to send a data packet it finds the route and establishes the connection. However, when the network becomes more complex, it is difficult to maintain the routing tables and consumes more power and bandwidth. Reactive scheme is based on on-demand routing, i.e., a satellite tries to find an optimal path to the destination only when there is a need to have an establishment of connection. Satellite-based networking has developed in
complexity over the years and numerous routing protocols have been proposed. Autonomous satellite systems must communicate and exchange routing information to make global routing possible. Border gateways run an exterior routing protocol that enables them to determine routes to other autonomous systems which are then propagated in the network through the internal routing protocol. The authors in~\cite{rnew6} proposed a new exterior gateway protocol called Border Gateway Protocol-Satellite version (BGP-S) that enables automated discovery of routes through the satellite network. For multi-layered satellite IP networks which includes GEO, LEO and MEO layers, a distributed multicast routing scheme is introduced in~\cite{rnew7}. The authors proposed a modification to the Multi-Layered Satellite Routing (MLSR) algorithm by adapting the algorithm to handle mobility of the satellites. It aims to reduce the cost of multicast trees rooted at the source.\\

Various protocols are associated with the network layer in order to maintain network connectivity. In \cite{Hi-DSN}, authors proposed several protocols for route discovery for a network of small satellites including, Neighbor Discovery Protocol, Network Synchronization Protocol, Decentralized Routing Protocol, Node Affiliation Protocol and Packet forwarding Protocol. The Neighbor Discover Protocol is investigated in detail in \cite{Hi-DSN}. The proposed protocol will enable the small satellites to advertise itself, find other satellites, and to achieve synchronization with other nodes within the transmission range. The neighbors possible for each satellite can be categorized into two types: new or re-occurring neighbors. A satellite performing neighbor discovery has no prior knowledge about a new satellite in terms of velocity, relative co-ordinates, frequency of transmission, etc. For a re-occurring neighbor, the satellite already knows all the information and is also synchronized with it. A node can establish neighbors by transmitting HELLO messages in an omni-directional way. Once a node acquires more and more neighbors, it can find new neighbors by sending HELLO messages using omni-directional antennas, but with null gain towards the established neighbors.  The HELLO burst reception is confirmed using FOUND\_YOU messages which further includes the assignment of codes and synchronization (SYNC) information. After exchanging the orbital parameters, the satellite ends the neighbor discovery process and starts transmitting data to the established neighbors using antenna arrays~\cite{Hi-DSN}. It is not necessary to have perfect synchronization for the neighbor discovery process at the beginning. The network synchronization protocol enables the satellites to achieve clock synchronization with respect to a reference satellite. The packet forwarding protocol helps to determine a satellite whether a packet needs to be forwarded, absorbed, or discarded.\\
\begin{table*}[!ht]
\caption{Routing techniques in small satellites}
\label{tab:routing schemes}
\renewcommand{\arraystretch}{1.3}

\begin{center}

\begin{tabular}{| >{\centering\arraybackslash}m{4.5cm} |  >{\centering\arraybackslash}m{3.5 cm} | m{5.5 cm}|}

\hline

\textbf{Routing Algorithm} & \textbf{Routing Metric} & \textbf{Advantages} \\

\hline

Handover optimized routing algorithm~\cite{routingalgorithms}& Connection matrix & Identifies the presence of inter-satellite links  \\

\hline
Bandwidth delay satellite routing~\cite{BDSR} & Delay and bandwidth & Ideal for LEO satellite networks \\
\hline
Destruction resistant routing algorithm~\cite{DRRA} & Link state & Survivability of the network is enhanced  \\
\hline
Steiner tree routing~\cite{Steinertree} & Number of hops & Limited overhead, supports a large number of satellites   \\
\hline
Distributed multi-path routing~\cite{distributedmultipathrouting} & N/A & Better end-to-end delay, instantaneous tracking of the changing topology of LEO satellite networks \\
\hline
Dynamic routing algorithm based on MANET~\cite{Manet} & N/A & Provides high autonomy, compatible functionality, limited overhead \\
\hline
\end{tabular}

\end{center}

\end{table*}

In~\cite{routingalgorithms}, various routing algorithms are discussed for LEO satellite systems. A handover optimized routing algorithm is proposed where the system model is based upon a constellation of 48 satellites called the Globalstar. The topology of the network of satellites at a particular time instant is called a topology slice, which keeps changing with time. The topology slice changes when a new inter-satellite link is added to the existing network or a link get broken in a space based network. Routing in such a dynamic environment is difficult. The connection state of each satellite with other satellites in the network is stored in a connection matrix. For the Globalstar constellation, the size of the connection matrix is 48x48, with each element representing whether inter-satellite communication exist between the satellites in a particular topology slice at a specific time instant. The authors in~\cite{BDSR} propose a Bandwidth Delay Satellite Routing (BDSR) which is based on optimization of both delay and bandwidth such that it balances the two performance indexes, satisfying the requirements of both bandwidth and timeliness in communication process. The routing strategy of BDSR is as follows: Suppose the source node \emph{$m_{1}$} needs to communicate with the destination node \emph{$m_{k}$} and there are \emph{n} reachable paths. The optimal path out of the \emph{n} paths must satisfy either minimum delay between \emph{$m_{1}$} and \emph{$m_{k}$} or maximum bandwidth between \emph{$m_{1}$} and \emph{$m_{k}$}. The authors did extensive simulations in NS2, the simulation environment includes 6 orbital planes with 11 satellites in each orbit at an altitude of 780 km above the Earth surface. They considered two scenarios, one in which the available bandwidth is constant and other scene with flexible bandwidth. The results indicate that when the bandwidth is fixed and taking only delay into account, the BDSR algorithm is reduced to shortest path algorithm. However, for flexible bandwidth case, on a link with the optimized delay performance, its bandwidth does not meet the requirements and it is observed that when the bandwidth of the link is best, delay is always increased by a large margin. It is concluded that this algorithm can adapt according to the link situations and then choose alternate paths, thus improving the overall system performance. A Destruction Resistant Routing Algorithm that is proposed in~\cite{DRRA} concentrates on avoiding invalid inter-satellite links and rerouting by selecting feasible paths in the network. The proposed algorithm uses off-line initialization strategy by computing the paths from each satellite to all other satellites in the network in advance. It uses two sub-procedures called cluster initiation and re-clustering to decrease algorithm complexity and to make sure that the reformed clusters follow the off-line cluster rules. The various rules of the proposed algorithm can be found in~\cite{DRRA}. Routing mechanism based on Steiner tree~\cite{Steinertree} and distributed multipath routing~\cite{distributedmultipathrouting} are the other two approaches for LEO satellite networks.  \\

The authors in \cite{Manet} introduce a new dynamic routing algorithm based on mobile ad-hoc network (MANET). Assuming the satellite network as a multi-hop wireless network, the whole network is divided into clusters to reduce the broadcast storm caused by change in network topology. This algorithm is established based on the assumption that intra-cluster satellite topology is known. Satellites in the network know which cluster other satellites belong to using the global node information table, and also it is assumed that all satellites know the intra-satellite cluster's relative locations and route to any other intra-satellite clusters with the help of routing table information. They emphasized Asynchronous Transfer Mode (ATM) based routing schemes for a network of small satellites by preparing virtual topology using virtual connections between the satellites. The proposed routing algorithm utilizes the advantages of both static and dynamic routing. Authors performed extensive simulations to analyze the adaptability of the proposed algorithm and proved that the algorithm provides the satellite network, high autonomy, compatible functionality, and low system overhead. The above discussions regarding the routing techniques in small satellites are summarized in Table~\ref{tab:routing schemes}.\\

Alternatively, technologies that are quite different from the protocols on which the Internet is built may be considered. Space communication may be subjected to intermittent connectivity such that there may not be at all times a continuous end-to-end path between the source and destination of data. It may also be subjected to long or variable signal propagation latency between satellites and ground stations. Such conditions can result in extremely long round-trip communication times between nodes in a satellite network, causing the TCP/IP communication protocols on which Internet communications are based to perform poorly. Over the past few years these considerations have led to the development of a ``Delay-Tolerant Networking" (DTN) architecture.\\

The DTN architecture was originally designed to enable automated network communication for space missions even in deep space, such as in relay operations between Mars landers and mission operations centers on Earth via spacecraft in orbit around Mars~\cite{Delaytolerent01}. Round-trip latencies in such missions may be as long as of tens of minutes. However, following development of the original architecture, potential applications in terrestrial networking, sparse sensor networks, and networks of Earth orbiters have emerged. The DTN architecture has evolved to address those cases as well~\cite{Delaytolerent02}.\\

The DTN architecture introduces an overlay network protocol termed ``Bundle Protocol" (BP), which utilizes protocols at the underlying ``convergence layer" to implement reliable transmission between BP nodes. Data issued via BP will be forwarded immediately by each node in the end-to-end path to the destination wherever possible, but where connectivity to the next node has temporarily lapsed the data will be retained in local node storage until communication is re-established. In contrast, in a network based on TCP/IP a transient partition in the network results in data being simply discarded.\\

Routing is also very different in a satellite network based on DTN. Since a satellite's approximate location at any time can be computed from its orbital elements and the locations of ground stations are fixed, opportunities for communication between satellites and ground stations can be anticipated and encoded in a ``contact plan" that can be uploaded to satellites. Communication among ground stations, over the terrestrial Internet, is at least potentially continuous in most cases. Taken together, these capabilities enable ``contact graph routing", the computation of efficient routes between satellites via ground stations over time-varying network topology. These routes may not be suitable for end-to-end conversational data exchange (VOIP) because satellites may at some times not be in contact with any ground stations. However, the store-and-forward nature of DTN communication enables these routes to be used effectively for non-conversational data exchange applications such as file transfer and asynchronous messaging.\\

The DTN architecture also includes mechanisms for data authentication and/or confidentiality.  Because data in a DTN-based network may reside in a node's local storage for minutes or hours while awaiting a future communication opportunity, these mechanisms are designed to secure information while it is at rest as well as in transit.\\

DTN has been demonstrated in a number of space flight contexts: the UK-Disaster Monitoring Constellation (2008), NASA JPL's Deep Impact Networking Experiment (2008), the International Space Station (2009-2013), and IntelSate-14 (2011). The DTN will enter continuous operational service on the International Space Station in 2015. It is possible that DTN may also be advantageous for small satellite missions~\cite{Delaytolerent04}. If small satellites used as relays were equipped with DTN technology, their relay functions would be only delayed, rather than interrupted, by lapses in radio contact. Simulations performed in the course of the study noted above indicate that the DTN architecture can increase small satellite relay data rates, and in general it has been proposed that DTN greatly improves communications performance in the presence of large propagation delay and link disruptions in a variety of satellite mission configurations~\cite{Delaytolerent05}. However, deploying DTN on small satellites characterized by limited processing speed, limited storage capacity, and power constraints may be challenging.

\section{Proposed Inter-Satellite Communication Solutions for Small Satellite Systems}
In this section, we present a few solutions to the challenges faced in implementing inter-satellite communications in small satellite systems. We have proposed solutions to some of the physical layer and data link layer challenges based on different areas of expertise in our research group.
\subsection{SDR Solution to Small Satellite Challenges} 
Today's wireless networks are characterized by a fixed network assignment policies which leads to inefficient utilization of the spectrum. Hence, a new communication paradigm is proposed referred to as NeXt Generation (xG) networks that utilizes Dynamic Spectrum Access (DSA) and Cognitive Radios (CR). Cognitive Radio aims to improve spectrum utilization by allowing unlicensed users to coexist with the  primary owners of spectrum (licensed users) without any interference to the communication. A ``Cognitive Radio" is formally defined as a radio that can change its
transmitter parameters based on interaction with the environment in which it operates~\cite{rnew1}. The two main characteristics of CR's are cognitive capability (referred to as the capability to sense the radio environment to identify the portions of spectrum that are unused at a particular time or location) and reconfigurability (enables the radio to be programmed dynamically depending on the radio environment). However, Cognitive Radio's impose severals challenges because of the fluctuating nature of the available spectrum which are explained in detail in~\cite{rnew1, rnew2}. The main
challenge in Cognitive Radio networks in a multi-hop/multi-spectrum environment is to integrate these functions in the layers of the protocol stack, in particular, network and transport layer, without any additional infrastructure
support which is investigated in~\cite{rnew4}. The authors emphasized on the distributed coordination between CR users through the establishment of a common control
channel. The authors also discussed current research challenges in terms of spectrum management functionalities such as cooperative spectrum sensing, cooperative spectrum leasing as well as
spectrum mobility in~\cite{rnew1, rnew2, rnew4,rnew8, rnew9}. The performance of CR's highly depend upon the activity of the primary users, in~\cite{rnew3} gives a detailed survey of the primary user radio activity model that have been used for cognitive radio networks. \\

The Software Defined Radio in general is a utilization of cognitive radios (CR) which is a system that implements all of their baseband functionalities in software. The term CR usually refers to secondary users in cognitive radio networks which concerns the problem of radio spectrum sharing, or detection of jamming that are not the cases in small satellites and hence SDR can be referred as a flexible radio to enable adaptive communication. This makes the SDR able to overcome hardware constraints imposed by standard hardware~\cite{sdrnew3}. In an SDR system, the Analog-to-Digital (ADC) and Digital-to-Analog converter (DAC) converts signals to and from the radio frequency front-end. The RF front end is used to down convert the signal to the lower frequency called an Intermediate Frequency (IF). The ADC will digitize signals and pass it to the baseband processor for further processes such as demodulation, channel coding, source coding, etc. Vulcan Wireless Inc. \cite{sdrnew4} has developed two SDRs optimized for usage in satellites. The first being the CubeSat SDR, which provides access to a wide variety of communication protocols and a data rate of up to 10 Mbps at S-Band~\cite{sdrnew4}. The second, MicroBlackbox Transponder, offers fewer protocols, and a lower data rate. These two systems support numerous S-Band frequencies (2-4 GHz) and work with a variety of communication protocols and encryption schemes~\cite{survey01}. However,  these SDRs do not support the Space Plug-and-Play Avionics (SPA) protocol for plug and play operation and does not use open source hardware or software. \\

In regards to an open architecture for the SDR in small satellites, Virginia Tech has made a new architecture available to solve this issue. The GNU radio architecture~\cite{sdr2} is an open-source initiative where the signal processing is carried out on GPP computers. GNU radio is adapted to the Universal Serial Radio Peripheral (USRP), which converts between base band and RF signals. The signal processing blocks are written in C++ and the graph is connected using the Python programming language.\\

The inter-satellite link allows the small satellites to communicate and exchange information with one another. ISL also allows the satellites to share resources to achieve the performance goal, while reducing the traffic load
to the ground. Software defined radio inter satellite links can provide relative position, time, and frequency synchronization for small satellites. An SDR inter-satellite link will be able to create automatically an ad-hoc inter-satellite link between the satellites and ground link capabilities. From the ground station one would be able to establish a network to cooperate and coordinate actions. High-speed data links of above 10 Mbps has been achieved, for instance, in the SWIFT SDR platform.\\
\subsubsection{SDR challenges in inter-satellite communications}
The SDR in small satellites offers the opportunity for cognitive and adaptive operation, multi-mode operation, radio reconfiguration, remote upgrade, as well as the potential to accommodate new applications and services without hardware changes. They also provide remarkable flexibility in dealing with bit rates, waveforms, and modulation and error correction schemes that can be supported by a single radio. While there are many advantages in the SDR payloads, they do face some challenges in small satellite payload applications. In~\cite{sdr1}, some of these challenges include:
\begin{enumerate}[label=(\alph*)]
\item Mass, power, and volume constraints for small satellites
\item Resource reservation required to make the SDR useful for potential update during a mission or reconfigurable for other missions
\item Bandwidth limit for remote software/firmware code update
\item Space radiation environment
\item What level of standardization should be adopted by the SDR
\end{enumerate}
The most widely used software architecture for SDR is the Software Communications Architecture (SCA). The SCA is an open architecture framework that tells designers how elements of hardware and software are to operate in harmony~\cite{sdr2}. The objective of this SDR software architecture is to introduce a transparency layer that decouples the waveform application from the underlying hardware, and to allow different objects to communicate with one another. This layer is known as the middleware. A political argument against SCA is that, it is not an open standard, as it is directly managed under the supervision of the Joint Program Executive Order.\\

In order to support the potential future functionalities and flexibilities, the SDR will require a certain amount of resources to be reserved. In~\cite{sdr1}, to make the SDR useful in supporting more complicated waveforms during a mission, a sizable memory and possibly CPU/FPGA processing power, and DC power capability need to be reserved at the beginning of the life of the mission. This translates to the increasing demands on the resources of the hosting satellite bus, i.e., size, mass, and power. The paper~\cite{sdrnew11} investigates the possibility to implement a new SDR architecture which utilizes a combination of Field Programmable Gate Array (FPGA) and field programmable RF transceiver to solve back-end and front-end challenges of a swarm of small satellites and thereby enabling reception of multiple signals using a single user equipment.\\

Security challenges play a major role in SDRs in small satellites because of the possibility of placing new software on the SDR unit through unauthorized and potentially malicious software installed on the platform~\cite{sdr3}, and the signals could easily be received by other hardware not in the network. Interferences coming from the external world as well are an issue. The security of this network could possibly be maintained by having cryptographic algorithms for instance. The cryptanalysis techniques developed later may render the current security evaluation insecure.  Another solution is through digital certification, which is a way of assuring that a public key is actually from the correct source. The Digital Certificate is digitally signed by a trusted third-party~\cite{sdr2}. Due to the digital certificate now being a signed data file itself, its authenticity can be determined by verifying its digital signature.\\

One of the current challenges is the responsiveness of the architecture of the SDR for small satellite implementation. Things to consider in the architecture of the SDR are: choosing a signal processor with a high precision reference oscillator and phase-locked loop for the master clock, an external interface, signal converter, intermediate frequency (IF) or in-phase and quad-phase (I/Q) baseband, optional IF up/down converter, buffer, low noise amplifier, RF up/down converter, a power amplifier, and antenna utilizing a transmit/receive switch, beam forming network, and transceiver. Larger satellites as compared to small satellites have bigger and advanced directional antennas. As the downlink data rates increase, using a low directivity space segment antenna is unsatisfactory and diminishes rapidly~\cite{sdr5}. The interference becomes an issue due to the increase of traffic within a fixed bandwidth causing the signal quality to degrade. Operationally, it is difficult to rely on a spacecraft’s attitude control system to maintain antenna pointing for a fixed beam antenna. Using an array of low directivity elements and steering
of the beam electronically proves to be a better solution.\\

The need of detailed zero-IF architecture for a triple-band VHF, UHF, and S band transceiver for multi-mode applications is proposed in~\cite{sdr6}. The VHF/UHF bands are chosen for the uplink/downlink, due to the feasibility and low cost to establish VHF/UHF ground stations. Also, the amount of ground facilities and amateur communities that can communicate in these bands are plentiful around the world, thus assisting to increase the communication window. The S-band will serve as the frequency band of the ISL for the small satellites to exchange data faster due to its ability in achieving high data rates. The goal for SDR is to move the digital domain (modulation/demodulation, encoding/decoding) as close as possible to the antenna, where the analog domain (band pass filters for frequency selection, low pass/ output filters for frequency conversion, and VGA for the gain control) reside. 

\subsubsection{Current implementations of SDR in small satellite systems}
\begin{figure}[h]
 \centering
 \renewcommand{\figurename}{Figure}
 \includegraphics[width =3in]{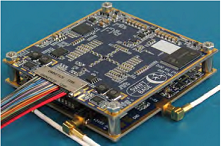}
 \caption{SWIFT RelNav SDR}
 \label{fig:sdr1}
\end{figure}

\begin{figure}[h]
 \centering
 \renewcommand{\figurename}{Figure}
 \includegraphics[width =3in]{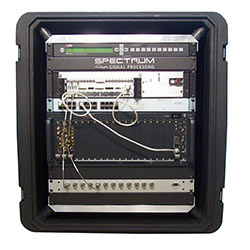}
 \caption{SDR-3000 Software Defined Radio platform}
 \label{fig:sdr2}
\end{figure}

Tethers Unlimited SWIFT-RelNav in~\cite{sdr_new1} is an SDR RF-based system that provides relative range and attitude determination capabilities as well as inter-satellite communications, shown in Figure \ref{fig:sdr1}. The SDR RelNav provides range sensing between satellites to better than 10 cm accuracy, inter-satellite crosslink data rates at 12 Mbps, bit error rates of 10\textsuperscript{-6}, and timing/frequency synchronization to better than 1 ns, 0.1 ppb. The SDR application in the ISL enables this system to perform ISL communication up to 10 km in range. This SDR RF based system proves to enable high data rates in satellites as well as operate in Ku and X bands. High data rate communications could eventually revolutionize space and science explorations. Figure 10 gives the description of the SWIFT-RelNav SDR. NASA is using the Spectrum SDR-3000 to enable satellites to communicate directly with one another for their Cross Link Integrated Development Environment (CLIDE) program. Utilizing these SDRs for NASA’s CLIDE project will enable NASA to develop inter-satellite cross links between satellites, enabling lower cost constellations of satellites to provide critical scientific data in a timely fashion~\cite{sdr_new2}. These direct satellite-to-satellite links allow for mesh connectivity and ad-hoc networking, thereby ensuring that a satellite communications network can provide full coverage of the earth. Multiple SDR-3000s will be used to simulate spacecraft in the lab and demonstrate full communication networking capabilities, including the inter satellite crosslinks. Figure \ref{fig:sdr2} shows the spectrum SDR 3000 that NASA is using for their test-bed for inter-satellite communications. \\

Using Software Defined Radio technology, we designed and implemented an optimal inter-satellite communications for a distributed wireless sensor network of small satellites~\cite{VTCconf}. The optimization of the ISL was achieved by designing a DS-CDMA communication using SDR.  The experimental result using the implemented system clarified our theoretical and simulated performances of the transmitted and received signals by their bit error rate measurements. For this research the physical layer and data link layer served as the focus of our work. For our SDR test-bed, the physical layer contained the USRP N210, which provides the transmission of raw bits over the antenna. The USRP N210 delivered a mode of operation from 0 to 6 GHZ and a transceiver which operated in the 400 MHz to 4.4 GHz range. We assumed a coherent system with transmission of bits modulated using BPSK and QPSK as well as exhibiting un-coded and convolutional coding techniques. The carrier frequency utilized was 2.4 GHz with AWGN, Rayleigh, and Rician Fading channel models in the case of Channel Side Information (CSI) being known at the receiver. For the data link, the multiple access technology types, Code Division Multiple Access (CDMA), Time Division Multiple Access (TDMA) and Frequency Division Multiple Access (FDMA) were considered in designing an optimum inter-satellite link. Direct Sequence Code Division Multiple Access was chosen due to factors such as multiple simultaneous transmission of signals, improved ranging accuracy from Global Navigation Satellite System (GNSS) technology and insensitivity to other satellites joining in and out the system. 

\subsection{Modular Antenna Array for Cubesats in Formation Flying Missions }
\begin{figure}[h]
 \centering
 \renewcommand{\figurename}{Figure}
 \includegraphics[width =3.2in]{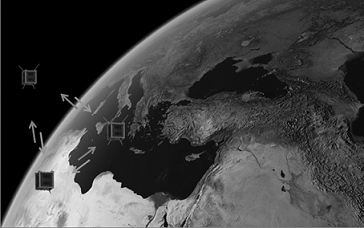}
 \caption{Illustration of a formation-flying mission concept with cubesats}
 \label{fig:antennadesign4}
\end{figure}

In this section, we provide an example of antenna design  for inter-satellite links for cubesat network. Figure~\ref{fig:antennadesign4} shows the concept of a distributed formation-flying cubesat mission where three satellites share information by means of inter-satellite links to cooperate and coordinate operations. Each cubesat makes use of an inter-satellite communication subsystem and reserves one of the cubesat face for the allocation of the inter-satellite link antenna. 

\subsubsection{Antenna specifications}
The antenna specifications presented in the Table ~\ref{tab:antennadesign1} are derived from the following general requirements:\\
\begin{enumerate}[label=(\alph*)]

\item Mission architecture - formation-flying mission\\
\item Platform - identical 1U cubesats\\
The antenna gain requirement has been derived from link budget figures with a receiver sensitivity of -100 dBm, and a transmit power of -33 dBm at 5.8 GHz for an inter-satellite range of 2 km. The selection of the operational frequency of 5.8 GHz is motivated by the use of an ISM band as well as the availability of COTS components. However, issues such as losses and manufacturing errors must be taken into account during the design phase. The minimum 10 dBi gain implies a maximum -3 dB beam width of 57 degrees which may not be enough to cover the exploration margin. Exploration margin provides information of the number of satellites in the formation that can be reached from a cubesat.
\end{enumerate}
\begin{table*}[t]
\caption{Specifications of antenna~\cite{antennadesign13}}
\label{tab:antennadesign1}
\begin{center}
\begin{tabular}{ | >{\centering\arraybackslash}m{2.5 cm} | >{\centering\arraybackslash}m{3.5 cm} | >{\centering\arraybackslash}m{2 cm} | >{\centering\arraybackslash}m{5.5 cm} | }
\hline
& \textbf{Parameter} & \textbf{Values} & \textbf{Comments}\\ \hline
\multirow{3}{*}{Physical} & Mass & 50 gm (max) & \\ \cline{2-4}
 & Thickness & 5 mm (max) & \\  \cline{2-4}
 & Size & 90x90 mm & \\ \hline
\multirow{7}{*}{Electrical} & Frequency & 5.8 GHz & ISM band \\ \cline{2-4}
 & Antenna gain & 10 dBi (min) & From inter-satellite link distance\\ \cline{2-4}
 & Exploration margin & $\pm$ 40 deg & From formation flying configuration \\ \cline{2-4}
 & Polarization & Circular & \\ \cline{2-4}
 & Return losses & $< $-10 dB & \\ \cline{2-4}
 & Input impedance & 50 $\Omega$ & \\ \cline{2-4}
 & Bandwidth & 1 MHz & \\\hline

\end{tabular}
\end{center}
\end{table*}

\subsubsection{Antenna array concepts}
\begin{figure}[h]
 \centering
 \renewcommand{\figurename}{Figure}
 \includegraphics[width =3in]{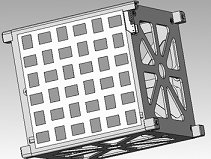}
 \caption{Antenna array attached to one of the cubesat faces~\cite{antennadesign12}}
 \label{fig:antennadesign5}
\end{figure}

\begin{figure}[h]
 \centering
 \renewcommand{\figurename}{Figure}
 \includegraphics[width =3.2in]{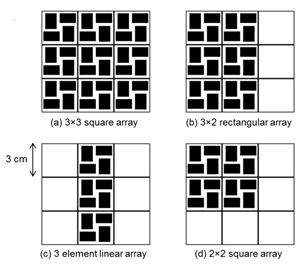}
 \caption{Antenna array configurations in a square of $9\times{9}$ cm}
 \label{fig:antennadesign6}
\end{figure}

From the above specifications, it is clear that in order to satisfy the exploration margin requirement, an antenna with electronic beam steering is required. We decided to choose a planar phased antenna array with a modular design where the available space for each element array is limited to $30\times{30}$ mm. From the mechanical requirements, the use of patch antennas as array elements is the most appropriate option at 5.8 GHz. Array elements are formed by a sub-array of 4 patches fed with sequential phase rotation in order to achieve the circular polarization. Thus, the maximum number of antenna elements in the planar array that fits in a cubesat side is nine under a 3x3 scheme. The antenna attached to the cubesat platform is depicted in Figure~\ref{fig:antennadesign5}.\\
Taking into account the array elements and available area to allocate the antenna, different array configurations are possible. The system engineer can select the most appropriate configuration depending on the required antenna aperture and exploration requirements, for example, linear, rectangular, and square arrays can be implemented. Linear geometries has beam steering in a single plane, whereas rectangular and square arrays can explore to any space direction in the exploration margin. Therefore, the proposed antenna array concept is modular and scalable and easy to manufacture as it is formed by identical sub-arrays. Figure~\ref{fig:antennadesign6} shows the different antenna array configurations possible in a square of 9x9 cm for small satellites.\\
\subsubsection{Antenna array functional description}
\begin{figure}[!h]
\centering\renewcommand{\figurename}{Figure}
\begin{subfigure}[t]{0.48\textwidth}
 \includegraphics[width =\textwidth]{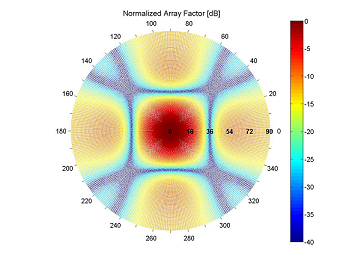}
 \caption{Normalized arrays factors: Broadside($\theta$ = $0^\circ$, $\psi$ = $0^\circ$)}
 \label{fig:antennadesign7a}
\end{subfigure}
\begin{subfigure}[t]{0.499\textwidth}
 \centering
 \renewcommand{\figurename}{Figure}
 \includegraphics[width = \textwidth]{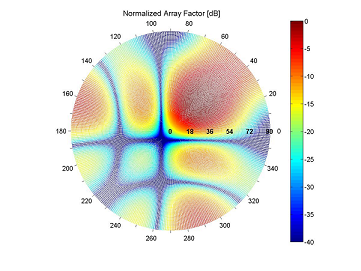}
 \caption{Normalized arrays factors: Broadside($\theta$ = $45^\circ$, $\psi$ = $45^\circ$)}
 \label{fig:antennadesign7b}
\end{subfigure}
\end{figure}

\begin{figure}[h]
 \centering
 \renewcommand{\figurename}{Figure}
 \includegraphics[height = 2.5in, width =5.2in]{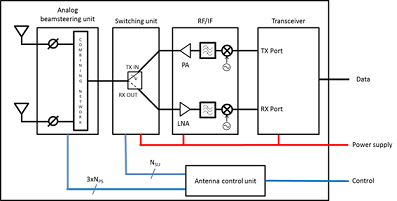}
 \caption{Antenna array block diagram}
 \label{fig:antennadesign8}
\end{figure}

Beam forming criteria can be computed from the relative position of the cubesats in the formation and the antenna beam is steered towards the preferred direction. Beam steering can also be used to modify the nominal direction to compensate deviations in the positions of the spacecraft in the formation. For beam steering, the complex weights must provide a progressive phase rotation in the antenna elements. The phase step depends on the array geometry and the direction to steer the beam. As shown in Figures~\ref{fig:antennadesign7a} and ~\ref{fig:antennadesign7b}, one digital phase shifter per antenna is used. The number of phase states in the phase shifter depends on the accuracy of the beam steering algorithm or pointing losses. Simulations results in MATLAB show that a digital control of $N_{PS}$= 3 bits to provide eight phase states (equivalent to a phase step of 45 degrees) are enough to fulfill the requirements with a pointing loss under 1 dB. The update rate of the phase shifters states depends on the attitude of the satellite and on the potential variation of the relative positions between cubesats.\\
Figure~\ref{fig:antennadesign8} shows the block diagram of the antenna subsystem for inter-satellite links for a TDD architecture with beam steering capabilities. Analog beam steering is performed by changing the phase shift of each antenna using a passive combining network that can be designed using miniaturized power splitters/combiners. Transmission and reception paths are separated by means of a miniaturized RF switch after the combining network (switching unit). An RF/IF stage has separate circuits for transmission and reception. Finally, the transceiver modulates/demodulates the incoming data/IF signal and interfaces with the satellite bus. 
Antenna control unit is responsible for calculating the phase shift between antenna elements for beam steering. This unit interface with OBDH (On-Board Data Handling) subsystem to receive information of the relative position between spacecraft, and on the other side it generates control signals for beam steering unit and RF switch.

\subsection{Optimum MAC Protocols for Inter-Satellite Communication for Small Satellite Systems}
In this section, we propose suitable MAC and routing protocols for a network of small satellites. The MAC protocol design plays a vital role in the performance of the system. It should consider numerous system parameters such as mission objective, network topology, number of satellites, etc. The MAC protocol must also take into account several system constraints, for example, limited on-board power and computing resources. \\

\subsubsection{Modified CSMA/CA/RTS/CTS protocol}
\begin{figure}[h]
\centering
\renewcommand{\figurename}{Figure}
\begin{subfigure}[b]{0.428\textwidth}
\includegraphics[width=\textwidth]{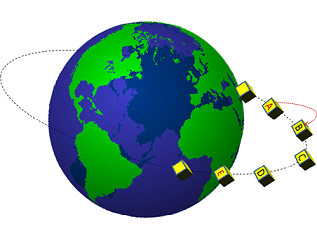}
\caption{Leader-follower formation flying pattern~\cite{ISC01}}
 \label{fig:routingprotocol01}
\end{subfigure}
\begin{subfigure}[b]{0.5\textwidth}
\centering\renewcommand{\figurename}{Figure}
\includegraphics[width=\textwidth]{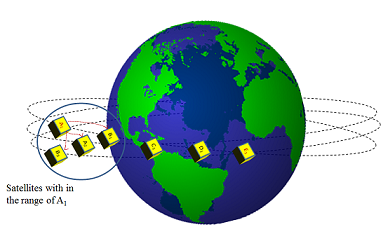}
\caption{Cluster formation flying pattern~\cite{ISC01}}
\label{fig:routingprotocol02}
\end{subfigure}
\end{figure}

We present a modified Carrier Sense Multiple Access, Collision Avoidance with Request-To-Send and Clear-To-Send protocol for a distributed network of small satellites which is based upon distributed coordination function, one of the services offered by the IEEE 802.11 standard. We proposed to use CSMA/CA/RTS/CTS protocol since it avoids hidden and exposed node problem compared to other traditional MAC protocols. A detailed explanation of the modified CSMA/CA/RTS/CTS protocol is given in~\cite{ISC08}. Our research mainly concentrates on three different small satellite configurations, namely, leader-follower, cluster, and constellation. Depending on the formation flying pattern, we proposed a reactive routing protocol which is based on on-demand routing, i.e., it establishes communication only when it is required. The data flow structure from the source satellite to the destination satellite for leader-follower, cluster, and constellation is shown in Figures~\ref{fig:routingprotocol01} and ~\ref{fig:routingprotocol02}.

For our system model, we consider 1U cube satellite with a transmission power of 500 mW to 2 W, operating at S-band frequency in the magnetic spectrum. We assume that the satellites are deployed in nearly circular lower Earth orbits. For leader-follower system, a single orbit is considered and for cluster, \emph{M} closely spaced orbits which are no wider than \emph{y} km are considered. For constellation configuration, we consider \emph{N} orbital planes, spaced \emph{x} degrees apart. For constellation formation flying pattern, it assumed that the satellites in distinct orbits join the network at different time instances in order to avoid collision at the poles. For all three small satellite configurations, it is assumed that all satellites share the same transmission frequency. The different system parameters used for simulation is given in Table ~\ref{tab:ISC_CSMA}.\\
\begin{table}[!h]
\begin{center}
\caption{Simulation Parameters}
\begin{tabular}{| >{\centering\arraybackslash}m{1.3in} | >{\centering\arraybackslash}m{1.3in} |}
\hline
\textbf{System Parameters} & \textbf{Value}\\ \hline
Size of cubesats & 1 U\\ \hline
 Transmission power & 500 mW to 2 W\\ \hline
 Orbital altitude & Lower Earth Orbit , 300 km\\ \hline
  Number of orbits - \emph{M, N} & 3\\ \hline
  Orbital separation, \emph{y} & 2 km\\ \hline
 Transmission frequency & 2.4 GHz \\ \hline
      Orbital velocity & 3 km per sec\\ \hline
      Inter-satellite range & 10 km to 25 km\\ \hline
      Number of packets simulated & 200 packets per satellite \\ \hline
   
 Data packet length & Exponential distribution\\ \hline
 Data packet arrival  & Poisson distribution \\ \hline
 DIFS  & 28$\mu$s\\  \hline
 SIFS  & 28$\mu$s\\  \hline
 RTS  & 50$\mu$s\\  \hline
 CTS  & 50$\mu$s\\  \hline
 ACK  & 14$\mu$s\\  \hline
 Average packet length & 1s\\ \hline
 Contention window size W & $2^{m}$\\ \hline
 \end{tabular}
 \label{tab:ISC_CSMA}
\end{center}
\end{table} 
 
\begin{figure}[!h]
\centering\renewcommand{\figurename}{Figure}
\begin{subfigure}[b]{0.4\textwidth}
\includegraphics[width=\textwidth]{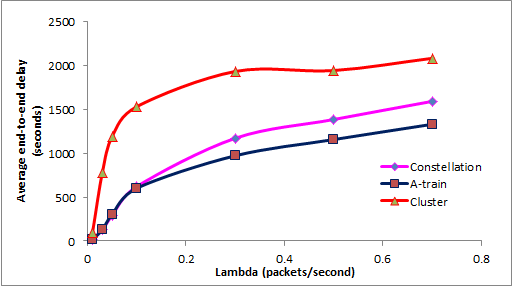}
\caption{Average end-to-end delay}
\label{fig:ISC_CSMA01}
\end{subfigure}
\begin{subfigure}[b]{0.4\textwidth}
\centering\renewcommand{\figurename}{Figure}
\includegraphics[width=\textwidth]{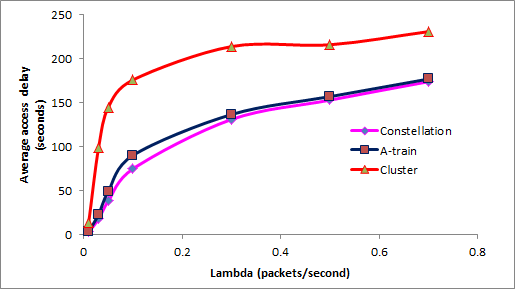}
\caption{Average access delay}
\label{fig:ISC_CSMA02}
\end{subfigure}
\begin{subfigure}[b]{0.4\textwidth}
\centering\renewcommand{\figurename}{Figure}
\includegraphics[width=\textwidth]{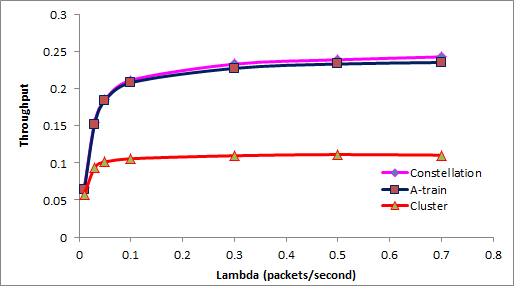}
\caption{Throughput}
\label{fig:ISC_CSMA03}
\end{subfigure}
\end{figure}

We did extensive simulations for the various formation flying patterns using an event driven simulator implemented in Java. The system performance was analyzed based on three different parameters, average end-to-end delay, average access delay, and throughput. The simulation results for the different configurations are given in detail in~\cite{ISC08}. In this paper, we present a brief review of the results already published in~\cite{ISC08}. Figures~\ref{fig:ISC_CSMA01},~\ref{fig:ISC_CSMA02}, and~\ref{fig:ISC_CSMA03} show the simulation results for the three different formation flying patterns. A scenario in which each of the configurations consists of 20 satellites per orbit is considered and it is assumed that the satellites are deployed at an altitude of 300 km above the Earth. \\
From Figures~\ref{fig:ISC_CSMA01} and~\ref{fig:ISC_CSMA02}, it can be observed that the average end-to-end delay and average access delay is more for cluster configuration compared to leader-follower and constellation formation flying patterns. This is due to the fact that, for cluster configuration,  all satellites share the same transmission frequency band, since there are more number of satellites within the range of each satellite results in more contention and thus causing increased delays. The leader-follower and constellation configurations have more throughput in comparison to cluster as shown in Figure~\ref{fig:ISC_CSMA03}. This is because the delay is more for cluster configuration and throughput is inversely related to delay. \\

We investigated the feasibility of the CSMA/CA/RTS/CTS protocol for various formation flying patterns of small satellites. The maximum throughput that can be achieved by using the proposed protocol for leader-follower and constellation formation flying pattern is around 24\%, and for cluster configuration is around 11\%. The major advantage of the proposed protocol is that it does not require strict synchronization between satellites. However, because of the large delays associated with this protocol, it is concluded that the proposed protocol is suitable only for missions that can tolerate communication delays, i.e., for missions that do not require near real time communications.\\ 

\subsubsection{Hybrid TDMA/CDMA protocol}

In this section, an overview of a novel hybrid TDMA/CDMA protocol for cluster of satellites is presented, which is explained in detail in~\cite{ISC07}. We suggested two different approaches, TDMA centric and CDMA centric, which will address the problem of multiple access in heterogeneous small satellite networks. A combination of TDMA with Direct Sequence CDMA (DS-CDMA) is investigated, where TDMA allows collision free transmission and DS-CDMA offers simultaneous transmission and better noise and anti-jam performance. \\

\begin{figure}[!h]
\centering\renewcommand{\figurename}{Figure}
\includegraphics[width =3.2in, height = 2in]{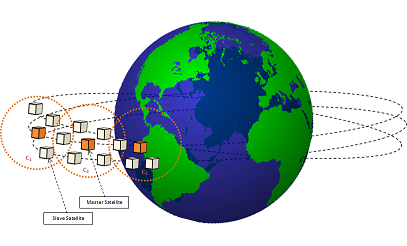}
\caption{Overlapped cluster of small satellites~\cite{ISC07}}
\label{fig:ISC_TDMACDMA01}
\end{figure}

Networking multiple spacecraft could be difficult since the space environment is dynamic and unpredictable with delayed or disrupted communication. Space communications also experience intermittent connectivity where it could be difficult to establish an end-to-end path between the source and destination satellite, and small satellite networks may have unexpected failures. Taking into account all these objectives, the small satellite network can be divided into clusters with each cluster having a master satellite and several slave satellites. The proposed system model is shown in Figure~\ref{fig:ISC_TDMACDMA01}. The slave satellites within a cluster communicates with the master satellite, and the master satellite forwards the data to the destination. If the member satellite needs to communicate with a satellite in another cluster, it first communicates with its own master satellite, which in turn communicates with the destination master satellite and thus forwards the data, thereby consuming a lot of power and hence, it is necessary to re-cluster the network. We propose to use closeness centrality algorithm for the selection of master satellite which satisfies the minimum power requirement (\emph{threshold}, $P_{th}$).\\

\begin{figure}[h]
\centering
\renewcommand{\figurename}{Figure}
\begin{subfigure}[t]{0.53\textwidth}
\includegraphics[width =3.3in, height = 2.5in]{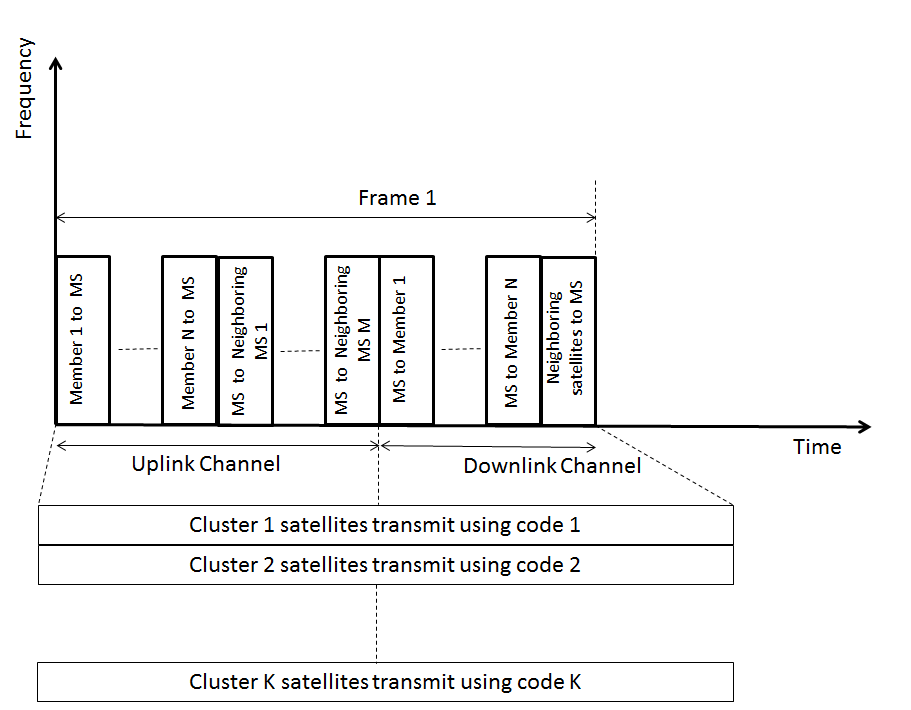}
\caption{The frame structure of hybrid TDMA/CDMA system (TDMA Centric)}
\label{fig:ISC_TDMAcentric}
\end{subfigure}
\begin{subfigure}[b]{0.428\textwidth}
\includegraphics[width =3.3in, height = 2.5in]{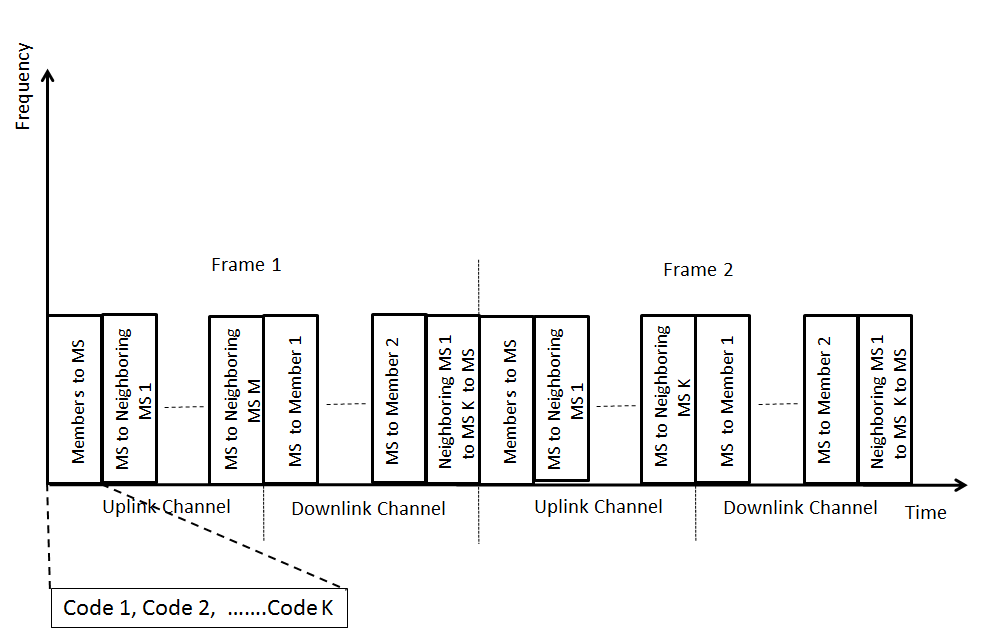}
\caption{The frame structure of hybrid TDMA/CDMA system (CDMA Centric)}
\label{fig:ISC_CDMAcentric}
\end{subfigure}
\end{figure}

The hybrid TDMA/CDMA can be implemented using two different approaches: TDMA centric and CDMA centric. In TDMA centric approach, each cluster is assigned a unique code. Each satellite has dedicated slots  for uplink and downlink to transmit the data to and from the master satellite. Multiple satellites from different clusters transmit in the same slot without interference using different codes. Figure~\ref{fig:ISC_TDMAcentric} shows the TDMA centric frame structure. In CDMA centric approach, each satellite is assigned a unique code. The member satellites can transmit data simultaneously to the master satellite in the first slot without interference using the respective orthogonal codes as shown in Figure~\ref{fig:ISC_CDMAcentric}. For the master satellite, there are dedicated slots to transmit data to the neighboring satellites and downlink slot for receiving data from the neighboring satellites. \\ 

\begin{figure}[!h]
 \centering\renewcommand{\figurename}{Figure}
 \includegraphics[width =3.2in, height = 2.5in]{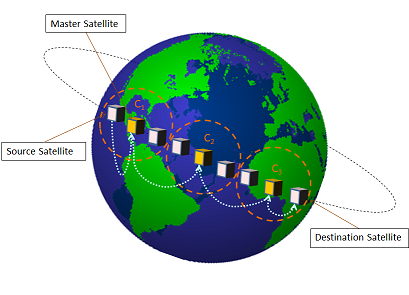}
 \caption{Simulation model~\cite{ISC07}}
 \label{fig:ISC_simulationmodel}
 \end{figure}

The performance of the hybrid TDMA/CDMA protocol is evaluated using three different parameters, throughput, average access delay, and average end-to-end delay respectively. For simplicity, the leader-follower formation flying pattern is chosen, with multiple satellites separated from each other at a specific distance and are placed in a single orbit. For simulation, the CDMA centric approach is used where we assumed a total of \emph{K} clusters, with \emph{N} satellites per cluster, and \emph{M} neighboring clusters. If a satellite has to transmit data, it first sends the information to the master satellite and the master satellite transmits the data to the destination through other master satellites. Figure~\ref{fig:ISC_simulationmodel} shows the proposed model and data flow structure from a source satellite to the destination satellite.    
 
\begin{figure}[!h]
\centering\renewcommand{\figurename}{Figure}
\begin{subfigure}[b]{0.4\textwidth}
\includegraphics[width=\textwidth]{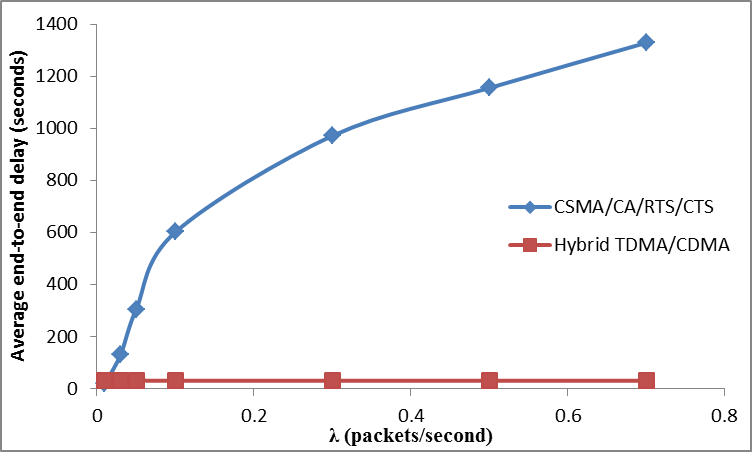}
\caption{Average end-to-end delay}
\label{fig:ISC_CDMAcentric_result02}
\end{subfigure}
\begin{subfigure}[b]{0.4\textwidth}
\centering\renewcommand{\figurename}{Figure}
\includegraphics[width=\textwidth]{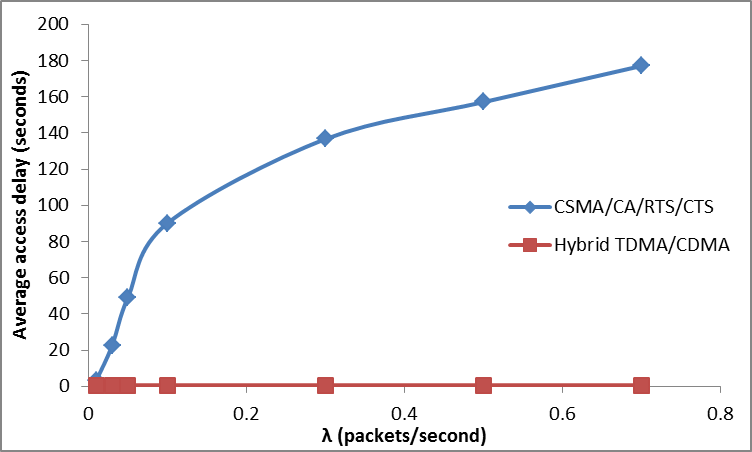}
\caption{Average access delay}
\label{fig:ISC_CDMAcentric_result01}
\end{subfigure}
\begin{subfigure}[b]{0.4\textwidth}
\centering\renewcommand{\figurename}{Figure}
\includegraphics[width=\textwidth]{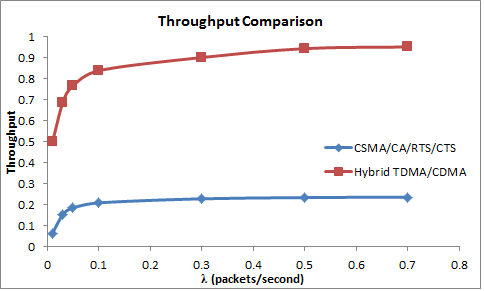}
\caption{Throughput}
\label{fig:ISC_CDMAcentric_result03}
\end{subfigure}
\end{figure}

 The parameters used for simulation are shown in Table~\ref{tab:ISC_CDMAcentric} which represent possible mission parameters our group may use in the future and are subject to change. 
 \begin{table*}[t]
 \begin{center}
 \caption{Simulation Parameters}
 \begin{tabular}{| >{\centering\arraybackslash}m{3in} | >{\centering\arraybackslash}m{3in} |}
 \hline
 \textbf{System Parameters} & \textbf{Value}\\ \hline
 Size of cubesats & 3 - 6 U\\ \hline
  Transmission power & 500 mW to 2 W\\ \hline
  orbital shape & Circular (for simplicity)\\ \hline
  Orbital altitude & Lower Earth Orbit , 300 km \\ \hline
   Number of orbits  & 1 (leader-follower)\\ \hline
   Number of satellites in each cluster & 3\\ \hline
\emph{N} (Number of slave satellites in each cluster) & 2\\ \hline
 \emph{K} (Number of clusters) & 3 - 9\\ \hline
\emph{M} (Neighboring clusters) & 2\\ \hline
      Transmission frequency & 2.4 GHz (ISM/S-band, Unlicensed band, higher throughput) \\ \hline
 Orbital velocity &	3 Km/s \\ \hline
 Inter-satellite range &	10 Km (from link budget analysis)\\ \hline
 Number of packets simulated &	10,000\\ \hline
 Packet arrival rate &	Poisson distribution\\ \hline
 Packet length & 	Exponential distribution\\ \hline
 Slot length &	100 ms\\ \hline
 Frame length &	0.6 s (6 slots/frame)\\ \hline
 
  \end{tabular}
  \label{tab:ISC_CDMAcentric}
 \end{center}
 \end{table*} 
We did extensive simulations using an event driven simulator implemented in Java. To obtain a reliable and stable result, the simulation runs consisted of 10,000 data packets. We assume that each satellite cannot generate a new message until all packets of the current message are transmitted, and data packets generated in the current frame have to wait for the next frame for transmission. Figures~\ref{fig:ISC_CDMAcentric_result01}, ~\ref{fig:ISC_CDMAcentric_result02},  and~\ref{fig:ISC_CDMAcentric_result03} show the average access delay, average end-to-end delay, and throughput of hybrid TDMA/CDMA protocol. We have also compared the results with the CSMA/CA/RTS/CTS protocol. The average access delay is almost constant for the hybrid protocol, around 0.6 seconds, since each data packet has to wait at least one frame long before it gets access to its allocated slot irrespective of the packet arrival rates. However, for the CSMA/CA/RTS/CTS protocol, the average access delay increases as the traffic increases due to network congestion. The average end-to-end delay is almost constant for the hybrid TDMA/CDMA protocol, but it increases for the CSMA/CA/RTS/CTS protocol for increasing traffic. The same logic applies here too as in the case of the average access delay. As the average access delay and end-to-end delay is inversely related to throughput, hybrid TDMA/CDMA protocol has a higher throughput of 95\% compared to the CSMA/CA/RTS/CTS protocol with a throughput of 24\% as shown in Figure~\ref{fig:ISC_CDMAcentric_result03}.\\
The proposed hybrid TDMA/CDMA protocol addresses the design needs of a large number of small satellites within a reconfigurable network. It allows for the simultaneous transmission of data in the allocated time slots by all satellites without interference. For a pure TDMA system, the addition of more satellites will be an issue  which can be overcome using CDMA technology based on clustering, thus supporting a large scalable network. The hybrid protocol has less delay compared to other MAC protocols, thereby making it suitable for missions that require tight communication links such as servicing and proximity operations. It can be implemented in two different frame structure: TDMA centric and CDMA centric. The TDMA centric hybrid protocol can be used in missions where the packet size varies considerably, where a variable number of slots (adaptive TDMA) are allocated depending on the size of the data packet provided there is a good control channel allocation. The cluster head must inform the members to refrain from using their slots in order to avoid collision. The CDMA centric system can be used when the packet size is relatively consistent and also for missions where it is required to broadcast some important information to the cluster members, for example, proximity operations.  
  
\section{Design Parameters for Inter-Satellite Communication Design Process}

\begin{table*}[t]
\renewcommand{\arraystretch}{1.3}
\caption{System design parameters for various applications}\label{sysdesign}
\centering
  \begin{tabular}{ | m {0.298cm}  m {0.298cm}  m {0.298cm}  m {0.298cm}  m {0.298cm} |m {3.2cm} | m {2cm} | m {1.9cm} | m {1.6cm} | m {1.9cm} |m {1.9
  cm}|}
     \hline
     \multicolumn{5}{|c|}{\textbf{OSI Layers}} & \textbf{System Design Parameters} & \textbf{Autonomous Operations}& \textbf{Earth Observation Missions} & \textbf{Deep Space Missions} & \textbf{Servicing or proximity operations} & \textbf{Distributed Processing}  \\  
     \multicolumn{5}{|c|}{(Potentially affected)} &  &  &  &  &  &   \\ \hline 
 \cellcolor{blue} A & \cellcolor{applegreen} T & \cellcolor{mediumcarmine} N & \cellcolor{mediumlavendermagenta} D & \cellcolor{selectiveyellow} P & Network topology (fixed/variable) & variable  &  variable/fixed  & variable & variable  & variable \\ \hline
    & & \cellcolor{mediumcarmine} N & \ & \cellcolor{selectiveyellow} P & Science data transmission frequency & low  &  high  & high/low & low  & high \\ \hline
    & & \cellcolor{mediumcarmine} N &  & \cellcolor{selectiveyellow} P & Navigation data transmission frequency & high  &  low & high & high  & high \\ \hline
    & & \cellcolor{mediumcarmine} N &  & \cellcolor{selectiveyellow} P & Command data transmission frequency & high  &  low & high & high  & high \\ \hline
    & & \cellcolor{mediumcarmine} N &  & \cellcolor{selectiveyellow} P &  Health and status data transmission frequency & low  &  low  & low & low  & low \\ \hline
  \cellcolor{blue} A & \cellcolor{applegreen} T & \cellcolor{mediumcarmine} N & \cellcolor{mediumlavendermagenta} D & \cellcolor{selectiveyellow} P & Power requirements & high  &  high  & high & high  & high \\ \hline
    & & & \cellcolor{mediumlavendermagenta} D & \cellcolor{selectiveyellow} P & Bandwidth requirements & high  &  high  & high & high  & high \\ \hline
    & & \cellcolor{mediumcarmine} N & \cellcolor{mediumlavendermagenta} D & \cellcolor{selectiveyellow} P & Real time access & high  &  low  & high/low & high  & high \\ \hline
   \cellcolor{blue} A & & & \cellcolor{mediumlavendermagenta} D & \cellcolor{selectiveyellow} P & Processing capabilities of each satellite & high  &  high/low  & high/low & high  & high \\ \hline
    \cellcolor{blue} A & & \cellcolor{mediumcarmine} N & \cellcolor{mediumlavendermagenta} D & \cellcolor{selectiveyellow} P & Reconfigurability & high  &  high/low  & high/low & high  & high \\ \hline
   \cellcolor{blue} A & & \cellcolor{mediumcarmine} N & \cellcolor{mediumlavendermagenta} D & \cellcolor{selectiveyellow} P & Scalability  & high  &  high/low  & high/low & high  & high \\ \hline
   \cellcolor{blue} A & \cellcolor{applegreen} T & \cellcolor{mediumcarmine} N & \cellcolor{mediumlavendermagenta} D & \cellcolor{selectiveyellow} P & Connectivity (intermittent/consistent)  & intermittent  &  consistent  & intermittent & intermittent  & intermittent/ consistent \\ \hline
   \cellcolor{blue} A & \cellcolor{applegreen} T & \cellcolor{mediumcarmine} N & \cellcolor{mediumlavendermagenta} D & \cellcolor{selectiveyellow} P & Variable data size  & low  & high  & high & low  & high \\ \hline
     \end{tabular}
\end{table*}  
The main drivers of ISC design process in general are the set of design parameters (constraints). They are obtained from the behavior of satellites operating in various types of constellations. 
The following are the design constraints from which specifications of one or more layers of the OSI framework are derived.
\begin{enumerate}[label=(\alph*)]
\item Network topology - Network topology is the arrangement of various elements (satellites, nodes in a computer network, sensor nodes, etc.) in a network. In a small satellite system, satellites can be arranged in a fixed or varying topology. 
 \item Frequency of data transmission - In distributed spacecraft systems there are four different data types that need to be exchanged between the satellites: science data, navigation data, spacecraft health/status data, and command/control data. The frequency of data exchange depends on the mission requirements. 
  \item Bandwidth requirements - The network of small satellites performing advanced functions requires high bandwidth, which largely depends on the mission and frequency of data transmission. 
  \item Real-time access - Extending networking to space will involve autonomous transfer of data without human intervention. There are various applications for small satellites such as, servicing or proximity operations, where data packets (involving time stamp information) need to transmit with a least amount of delay. Satellites need to have real-time access to the communication channel for such applications. 
  \item Processing capabilities of each satellite - Depending on the mission, each small satellite will have distinct processing capabilities. For a centralized system, the mother satellite in the system would have higher processing capabilities in comparison to daughter satellites. Daughter satellites can transmit raw data to the mother satellite, which in turn process the data, reduce the size, execute necessary error correction techniques, and transmit it to the ground station. For a purely distributed network, the processing capability of each satellite in the system would be comparable.
  \item Reconfigurability and scalability - The two important requirements of small satellite sensor networks are reconfigurability and scalability. Applications and protocols implemented in these networks should check for node failures or addition of new nodes, and reconfigure itself to maintain mission objectives. The various layers of the OSI model should be designed to support different network architectures, control over network topology, and also assist high degree of scalability.
\item Connectivity - The challenging space environment and node mobility will cause the low power small satellites to periodically lose connection with each other. Networking under such intermittent connectivity is demanding, as many of the terrestrial protocols are not suitable in this context. Thus, their performance deteriorates drastically as connectivity becomes intermittent and short-lived. Hence, routing is one of the biggest problems to overcome. The existing terrestrial protocols need to be modified in order to meet the requirements in space applications.    
\item Variable data size - The data size can vary considerably from several kilobits to megabits depending on mission applications. The protocols should be designed such that they are capable of adapting based on the size of data.
 \end{enumerate} 
 
System design parameters (constraints) are dependent on mission types leading to different applications, such as autonomous operations \cite{TanyaAgent}, Earth observation missions, deep space missions, servicing or proximity operations \cite{proximityref}, and distributed processing \cite{distributedprocessingref}. For example, autonomous operations require variable network topology, science data, health and status data need to be transmitted less frequently, but frequency of navigation data would be very high and data size can be variable. For missions demanding autonomous functionalities, small satellites would require high power, bandwidth, real time access to the channel, and processing capabilities. These type of networks would experience intermittent connectivity and the topology would be highly dynamic in nature. Design processes should capture this information and pass it to the OSI framework ensuring consistent and reliable ISC among satellites.\\

Table \ref{sysdesign} illustrates the criticality of the various system design parameters depending on the different applications of small satellites. The first column of the table is color-coded (based on Figure \ref{fig:osi}) to show the relationship between the design parameters (constraints) and various OSI layers at a specific level of abstraction in the design process as an example.\\

For example, network topology can be fixed or variable depending on the mission requirements. Hence, the various design parameters of the OSI model are potentially affected. The algorithms and software programs designed in the application layer should incorporate the change in network topology. Considering the dynamic topology, the transport and network layer parameters must choose the optimum routing metric such that highest performance can be achieved by minimizing the delay. Depending on the change in topology, the MAC protocols must be designed to ensure fairness among different satellites in the system, which in turn affects the physical layer parameters. The network, and physical layer parameters are primarily affected by the rate at which various data (science, navigation, command and health/status) are transmitted among the small satellites. Depending on the frequency of data transmissions, network layer must choose ideal routing metric and routing path. The frequency of data transmissions predominantly influence all physical layer parameters including bandwidth, data rate, antenna design parameters, transmission frequency, etc. \\

This relationship may represent ``derived from", ``verify", etc.  However, the table does not present all the relationships. The table should be understood with the disclaimer that the design parameters in column 2 have varying degrees of impact on the OSI layers mentioned in column 1.
\section{Future Research Directions}
Extending networking to space will involve large number of satellites with dynamic topology requiring high data rate communication. It is important to develop robust ad-hoc networking of mobile elements to co-ordinate timing, position, and spacing among the satellites with advanced methods of channel accessing and routing schemes. Hence it is necessary to understand the relation between different functionalities of the OSI model and interdependency of the various parameters. Cross layer optimization allows communication between layers by permitting one layer to access the data of another layer to exchange information and enable interaction. To this end, there are many
questions to be addressed. For example, can we implement the associated mission as a satellite sensor network? If so, how can we modify the existing OSI layer design such that it can support real-time and high-rate communications with extremely high reliability and security? Apparently, how are we going to deal with the complexity of the system? The research on inter-satellite communication is still in its early stage, and its the impact would be significant. Therefore, Cross-layer optimization for small satellites represents another research area to be investigated.\\

Future missions will demand autonomous transfer of data where today such transfers involve high levels of manual scheduling from Earth.  To solve these issues, new agent based computing platforms are proposed i.e., the satellites should have capabilities to perform intelligent improvements based on the situations. For example, each satellite or agent in the system receives information from the neighboring satellites and decides the actions it should perform. Satellites need to discover the current network topology they have formed and should determine whether that situation is appropriate to initiate communication. In other words, satellites should recognize all possible combinations of network topologies they may form and wisely decide a suitable one for communication, so that, an optimum system performance can be achieved. Future research is expected to analyze the performance of different protocols and algorithms for large number of satellites for highly autonomous systems. It would be worth to investigate on developing reconfigurable architecture and software algorithms for such agent based systems to achieve higher levels of autonomy.
\section{Conclusions}
Advancements in communication and navigation technology will allow future missions to implement new and more capable science instruments, greatly enhance missions within and beyond Earth's orbit, and enable entirely new mission concepts using a large number of affordable spacecraft. Development of novel and efficient wireless technologies for inter-satellite communications are essential for building future heterogeneous space networks to support a wide range of mission types and to meet the ever-increasing demands for higher data rates with minimal latency. As new mission concepts are developed, and human exploration intensifies, communication among heterogeneous platforms is challenging. There have been significant research efforts in the area of inter-satellite communications in small satellite systems which is presented in this paper. We conducted a detailed study on the various design issues based on the OSI model, with main focus in the last three layers. \\

Physical layer parameters such as modulation, coding, link design, antenna design, and the use of software defined radio has been investigated and a detailed description of these research efforts is provided in our survey. A detailed study of different MAC protocols suggested for small satellite networks has been presented. The various MAC protocols are compared with respect to topology, synchronization, advantages and disadvantages. In the topology classification, we differentiate the protocols as centralized and distributed schemes. On the other hand, synchronization is required in most contention-free and hybrid protocols. Different routing schemes used in small satellite networks are also presented. Earth based inter-networking technologies cannot be implemented in space because of the dynamic and unpredictable nature of the space environment. These issues can be overcome using Disruptive Tolerant Networking (DTN) protocols which is also described in detail in this paper. The DTN technology development will enable future networking capabilities through out the solar system. \\

We also demonstrated some of the solutions for the challenges faced by the small satellite systems. This includes implementation of software defined radio for small satellite systems, designing a modular antenna array for cubesats flying in formation, and developing feasible multiple access protocols for inter-satellite communications in small satellite systems. Some of the proposed or already launched missions involving formation flying concept is also discussed. Lastly, we provided a set of design parameters that need to be considered while designing and building multiple satellite missions involving inter-satellite communications. This survey will serve as a valuable resource for understanding the current research contributions in the growing
area of inter-satellite communications and prompt further
research efforts in the design of future heterogeneous space missions. 

\section{Acknowledgments}

The authors thank their colleagues \textbf{Dr. Qing-An Zeng}, \textbf{Dr. Sun Yi}, \textbf{Rawfin Zaman} and \textbf{Solomon Gebreyohannes} who provided insight and expertise that greatly assisted the study and preparation of this research paper.\\

This work was funded through the Langley Professor grant from the National Institute of Aerospace. This work was partially supported by the North Carolina Space Grant under New Investigator Award No. NNX10A168H. The research described in this paper was performed in part at the Jet Propulsion Laboratory, California Institute of Technology, under a contract with the National Aeronautics and Space Administration.

\bibliographystyle{IEEEtran}
\bibliography{surveyreferences02}

\end{document}